\documentclass[%
 reprint,
 amsmath,amssymb,
 aps,
]{revtex4-1}
\usepackage{color}   
\usepackage{graphicx}% Include figure files
\usepackage{dcolumn}% Align table columns on decimal point
\usepackage{bm}% bold math
\usepackage{amsfonts,amsthm,amsmath,amssymb,graphicx}
\usepackage{braket}
\usepackage{hyperref}% add hypertext capabilities
\begin{document}

%\preprint{APS/123-QED}

\title{Holographic entanglement contour, bit threads, and the entanglement tsunami}% Force line breaks with \\
\author{Jonah Kudler-Flam}
\email{jkudlerflam@uchicago.edu}
%  \altaffiliation[Also at ]{Physics Department, XYZ University.}%Lines break automatically or can be forced with \\
\author{Ian MacCormack}%
 \email{imaccormack@uchicago.edu}
\author{Shinsei Ryu}%
 \email{ryuu@uchicago.edu}
\affiliation{%
  Kadanoff Center for Theoretical Physics,
  University of Chicago, Illinois 60637, USA.
}%
% \author{Ann Author}
%  \altaffiliation[Also at ]{Physics Department, XYZ University.}%Lines break automatically or can be forced with \\
%  \email{Second.Author@institution.edu}
% \affiliation{%
%  Authors' institution and/or address\\
%  This line break forced with \textbackslash\textbackslash
% }%

% \collaboration{MUSO Collaboration}%\noaffiliation

%  \homepage{http://www.Second.institution.edu/~Charlie.Author}
% \affiliation{
%  Second institution and/or address\\
%  This line break forced% with \\
% }%
% \affiliation{
%  Third institution, the second for Charlie Author
% \affiliation{%
%  Authors' institution and/or address\\
%  This line break forced with \textbackslash\textbackslash
% }%

% \collaboration{CLEO Collaboration}%\noaffiliation

\date{\today}% It is always \today, today,

\begin{abstract}
  We study the entanglement contour,
  a quasi-local measure of entanglement, and propose a generic formula for the contour in 1+1d quantum systems.
  We use this formalism to investigate the real space entanglement structure of
  various static CFTs as well as local and global quantum quenches.
  The global quench elucidates the spatial distribution of entanglement entropy
  in strongly interacting CFTs
  and clarifies the interpretation of the entanglement tsunami picture.
  The entanglement tsunami effectively characterizes the non-local growth of entanglement entropy
  while the contour characterizes the local propagation of entanglement.
  We generalize the formula for the entanglement contour to arbitrary dimensions
  and entangling surface geometries using bit threads,
  and are able to realize a holographic contour for logarithmic negativity
  and the entanglement of purification by restricting the bulk spacetime to the entanglement wedge.
  Furthermore, we explore the connections between the entanglement contour, bit threads, and entanglement density in kinematic space. 
% \begin{description}
% Secondary publications and information retrieval purposes.
% \item[PACS numbers]
% May be entered using the \verb+\pacs{#1}+ command.
% You may use the \texttt{description} environment to structure your abstract;
% use the optional argument of the \verb+\item+ command to give the category of each item. 
% \end{description}
\end{abstract}

% \pacs{Valid PACS appear here}% PACS, the Physics and Astronomy
%                              % Classification Scheme.
% %\keywords{Suggested keywords}%Use showkeys class option if keyword
\maketitle
\tableofcontents

\section{Introduction}
Many-body entanglement has played a central role in modern theoretical physics. It has been used for characterizing quantum critical systems \cite{2003PhRvL..90v7902V} and topological order \cite{2006PhRvL..96k0405L,PhysRevLett.96.110404} and is a key ingredient for the emergence of bulk spacetime in gauge-gravity duality \cite{2006PhRvL..96r1602R, 2006JHEP...08..045R, 2010GReGr..42.2323V}. Furthermore, it is a useful probe of thermalization in many-body systems. The standard measure for pure state entanglement is the von Neumann entropy, which is a highly nonlocal quantity associated to an entire codimension-1 region of spacetime. Likewise, its holographic dual is the area, $\mathcal{A}$, of the extremal surface in the bulk that is homologous to the boundary subregion of interest \cite{2006PhRvL..96r1602R, 2006JHEP...08..045R,2007JHEP...07..062H}
\begin{align}
    S_{vN} = \frac{\mathcal{A}}{4 G_N}.
    \label{RT}
\end{align}
It is desirable to decompose the von Neumann entropy into local quantities that describe the contribution of each local degree of freedom in a given subsystem to the total entanglement. This would act as a quantitative measure complementing the geometric picture of real space entanglement seen in tensor networks. For instance, a suitable fine-grained measure would show the dominance of the entropy for ground states of gapped Hamiltonians near the entangling surface, and a constant entropy density for thermal states. Progress on decomposing entropy into a fine-grained quantity was made with the introduction of the entanglement contour \cite{2014JSMTE..10..011C}\footnote{We note that a similar construction was studied earlier in Ref. \cite{2004PhRvA..70e2329B}.}.

The entanglement contour at a point $x$ in subregion $A$, $s_A(x)$, is non-uniquely defined through five conditions:
\begin{enumerate}
    \item Positivity: $s_A(x) \geq 0 \quad \forall  \quad x \in A$.
    \item Normalization: $\int_A s_A(x) d^dx = S(A)$ where $S(A)$ is the von Neumann entropy of $A$.
    \item Invariance under spatial symmetry transformations: If $T$ is a symmetry of the reduced density matrix, $\rho_A$, that exchanges two sites $i,j \in A$, then $s_A(i)= s_A(j)$.
    \item Invariance under local unitary transformations: 
    If $\rho'_A = U_X \rho_A U_X^{\dagger}$, where $U_X$ is a local unitary supported on $X \subseteq A$, then $s_A(X)$ is equal for both $\rho_A$ and $\rho_A'$. Here,
    \begin{align}
        s_A(X) = \int_X s_A(x) d^dx.
    \end{align}
    \item Upper bound: If $ \mathcal{H}_A = \mathcal{H}_{\Omega} \otimes \mathcal{H}_{\bar{\Omega}}$ and $\mathcal{H}_X$ is contained within $\mathcal{H}_{\Omega}$, then
    \begin{gather}
        s_A(X) \leq S(\Omega).
    \end{gather}
\end{enumerate}
It is an open program where more conditions may need to be defined. Contour functions have previously been constructed for free fermions, harmonic lattices, single intervals in 1+1d holographic CFTs, and 1+1d inhomogeneous critical systems \cite{2014JSMTE..10..011C,2017JPhA...50E4001C,2018arXiv180305552W,2018arXiv180704179A,2018JSMTE..04.3105T}, with each having a unique construction. We find that bit threads \cite{2017CMaPh.352..407F}, an alternative description of holographic entanglement entropy, elucidate the non-uniqueness of the entanglement contour\footnote{We also thank Erik Tonni for directing us to his slides from It from Qubit Bariloche where a connection between bit threads and the entanglement contour was originally discussed.}.

We organize the paper as follows: In section \ref{1+1}, we unify the entanglement contour for generic 1+1d systems and compute the contour for a range of static systems with distinct entanglement structures, including thermal systems and defect CFTs. In section \ref{bitthread}, we use bit threads to propose a holographic realization of the entanglement contour for any dimension and entangling surface geometry. In section \ref{kinematic}, we tie together the notions of the entanglement contour, entanglement density, and bit threads with kinematic space. We then introduce dynamics in section \ref{quench} with both local and global quantum quenches. The contour proves to be particularly useful in characterizing non-equilibrium dynamics and clarifies the interpretation of the entanglement tsunami picture in ergodic CFTs \cite{2014PhRvL.112a1601L,2014PhRvD..89f6012L,2016JHEP...07..077C,2015PhRvD..92l6004L}. 

\section{Entanglement contour for 1+1d systems}
\label{1+1}

In order to construct an entanglement contour
for generic quantum systems, 
let us partition a subregion $A$ into $n$ degrees of freedom $\{A_i\}$. The entropy of $A$ can be expanded in terms of conditional entropies
\begin{align}
  S(A)
  &= \frac{1}{2} \sum_{i=1}^n
    \Big[
    S(A_i| A_{1}\cup \dots \cup A_{i-1})
    \nonumber \\
  &\qquad \quad
    +S(A_i| A_{i+1} \cup \dots \cup A_n) \Big],
    \label{contourDef}
\end{align}
where conditional entropy is defined as
\begin{align}
    S(A|B) =S(A\cup B) -  S(B).
\end{align}
For conciseness, we replace the first and last terms of (\ref{contourDef}), $S(A_1|A_1)$ and $S(A_n|A_n)$, with $S(A_1)$ and $S(A_n)$, respectively. This leads us to a natural entanglement contour
\begin{align}
  s_A(A_i)
  &= \frac{1}{2}
    \Big[ S(A_i| A_{1}\cup \dots \cup A_{i-1})
    \nonumber \\
  &\qquad \quad
    +S(A_i| A_{i+1} \cup \dots \cup A_n)\Big].
    \label{contour_eq}
\end{align}
We stress that this choice of a contour function is not the only function that may satisfy the requirements. Technically, this is an entanglement contour for any the dimension, though for higher dimensions there will be an ambiguity regarding the ordering of $\{A_i \}$. Therefore, we restrict our current focus to 1+1d systems and return to higher dimensions in Section \ref{bitthread}. In \cite{2018arXiv180305552W}, an entanglement contour was proposed for single intervals in 1+1d holographic CFTs
\begin{align}
    s_{A}(A_2) = \frac{1}{2}\left[S(A_1 \cup A_2) + S(A_2 \cup A_3) - S(A_1) - S(A_3) \right],
    \label{wen}
\end{align}
where $A_1 \cup A_2 \cup A_3 = A$. This was derived using natural slicings of the Ryu-Takayanagi surface arising from the bulk extension of the modular flow. (\ref{contour_eq}) generalizes this formula to multiple intervals and generic quantum systems. 
We will now prove that (\ref{contour_eq}) satisfies all five conditions for an entanglement contour completely generally, without any reference to holography or quantum field theory.
\begin{enumerate}
    \item Positivity: This follows directly from the strong subadditivity of von Neumann entropy \cite{araki1970}.

\item{Normalization:}
By construction, normalization is satisfied.

\item{Symmetry:}
We take a spatial symmetry that exchanges $i,j \in A$, $T \rho^{A} T^{\dagger} = \rho^{A}$. Each component of (\ref{contour_eq}) is invariant under such a transformation, so $s_A(i) = s_A(j)$.

\item{Invariance under local unitary transformations:}
As stated in \cite{2018arXiv180305552W}, the causal property of entanglement entropy ensures that all components of (\ref{contour_eq}) are stationary under unitaries acting nontrivially only on $A_i$. 

\item{Upper bound:}
We take a subregion $A_2 \subset A$ that is contained in a factor space $\Omega_A$. We must then prove that $s_A(A_2) \leq S(\Omega_A)$. By the additivity of the contour and subadditivity of von Neumann entropy, 
\begin{align}
    s_A(A_2) \leq s_A(\Omega_A) \leq S(\Omega_A).
\end{align}
\end{enumerate}

\subsection{Static examples}
We reproduce results derived in \cite{2018arXiv180305552W, 2018arXiv181011756W}, so that we can refer to them later on. For the vacuum, the contour for the 1+1d subinterval, $(-l/2, l/2)$, is
\begin{align}
    s_{A}(x) = \frac{c}{6} \left(\frac{l}{\frac{l^2}{4} - x^2} \right)
\end{align}
and for a thermal state with inverse temperature $\beta$,
\begin{align}
      s_A(x) = \frac{c{\pi}}{{6 \beta}} \left(\coth\left(\frac{\pi (x + \frac{l}{2})}{\beta} \right)+\coth\left(\frac{\pi (\frac{l}{2}-x)}{\beta} \right) \right).
      \label{therm_cont}
\end{align}
It was also shown in \cite{2018arXiv181011756W} that for the warped CFT dual to ${\it AdS}_3$ with chiral boundary conditions \cite{2013JHEP...05..152C}, 
\begin{align}
    s_{A}(x) = \frac{c}{12} \left(1 + \frac{l}{\frac{l^2}{4} - x^2} \right).
\end{align}
Interestingly, this is identical to the regular CFT result with an additional ``thermal" term.

\subsubsection{Defect CFTs}
Using a simplified version of the Randall-Sundrum model, we can model the
holographic dual of a  defect CFT (dCFT)
as two copies of ${\it AdS}_3$ with a deficit angle \cite{2011PhRvL.107j1602T,2008JHEP...03..054A}
with the metric
\begin{align}
    ds^2= d\rho^2 + \cosh^2 \frac{\rho}{l_{AdS}} \left( \frac{dy^2 -dt^2}{y^2} \right).
\end{align}
Above, $l_{AdS}$ is the ${\it AdS}$ curvature radius, $\rho$ is the radial coordinate (with asymptotic boundaries at $\rho= \pm \infty$), and $y$ is the transverse spatial coordinate. The two copies of the deficit angle ${\it AdS}$ are glued together at a tensionful brane, which models the defect. For a brane of tension $\lambda$, the radial coordinate $\rho$ ranges between $-\infty$ and
\begin{align}
    \rho_*= l_{AdS}\tan^{-1}(l_{AdS}\lambda).
\end{align} To consider the dCFT, we double the domain by attaching $\rho \in ( -\rho_*,\infty)$. Using (\ref{contour_eq}), we find
\begin{align}
    s_{A}(x) = \frac{c}{6} \left(\frac{l}{\frac{l^2}{4}- x^2} +  \rho_*\delta(x) \right).
\end{align}
Naturally, all of the boundary entropy is localized on the defect brane.

\subsubsection{Black hole microstates}
Black hole microstates are dual to high-energy eigenstates of the CFT, which can be formed by the operator-state mapping of local heavy operators, $\psi$, with conformal dimensions $h_{\psi} = \bar{h}_{\psi}$. The entanglement entropy is calculated by the four-point function
\begin{align}
    S_{\psi}(x_1, x_2) = \lim_{n \rightarrow 1} \frac{1}{1-n} \log \bra{\psi} \sigma_n(x_1) \bar{\sigma}_n(x_2) \ket{\psi}.
\end{align}
Leveraging progress in the calculation of conformal blocks for ``heavy-heavy-light-light" correlation functions \cite{2014JHEP...08..145F}, the entropy is found to be \cite{2015JHEP...02..171A,2015JHEP...01..102C}
\begin{align}
    S_A = \frac{c}{3}\log \left(\frac{\beta_{\psi}}{\pi \epsilon}\sinh\left( \frac{l \pi}{\beta_{\psi}}\right) \right),
\end{align}
where $l$ is the length of the interval and $\beta_{\psi}$ is an effective temperature
\begin{align}
    \beta_{\psi} = \frac{2\pi}{\sqrt{24h_{\psi}/c -1}}.
\end{align}
We then find the entanglement contour for the high-energy eigenstate
\begin{align}
    s_A(x) = \frac{c {\pi}}{{6 \beta_{\psi}}} \left(\coth\left(\frac{\pi x}{\beta_{\psi}} \right)+\coth\left(\frac{\pi (l-x)}{\beta_{\psi}} \right) \right),
\end{align}
analogous to a truly thermal state. While these states clearly obey the Eigenstate Thermalization Hypothesis \cite{PhysRevE.50.888}, the contour may be particularly useful for understanding systems whose eigenstates do not thermalize, e.g. many-body localized phases and scar states \cite{2018NatPh..14..745T}.

\subsubsection{Massive deformation}

We perturb our CFT by a massive deformation as was done in \cite{2006PhRvL..96r1602R,2006JHEP...08..045R}. This is crudely done by capping off the bulk IR geometry with characteristic correlation length $\xi$. For small intervals, we still have the critical vacuum entropy
\begin{align}
    S = \frac{c}{3}\log \frac{l}{a},
\end{align}
but for large intervals,
\begin{align}
    S = \frac{c}{3}\log \frac{\xi}{a}.
\end{align}
In order to probe the IR region, we work with a large interval $(-l/2, l/2)$ with $l \gg \xi$. It is then clear that the contour is trivial away from the entangling surfaces, and is universal and independent of $l$ near the entangling surfaces
\begin{align}
    s_{A} (x) = \begin{cases} 
      \displaystyle
    \frac{c}{6(x-\frac{l}{2} )} & -\frac{l}{2} <x <-\frac{l}{2}+\xi \\ \\
      \displaystyle
    0 &  -\frac{l}{2} +\xi <x <\frac{l}{2}- \xi \\ \\
      \displaystyle
    \frac{c}{6(\frac{l}{2}-x)} & \frac{l}{2}- \xi <x <\frac{l}{2}
    \end{cases}.
\end{align}
This is the area law behavior we would hope for the contour to capture, though the polynomial decay of the contour is most likely due to our crude setup. We expect that more realistic gapped theories will have exponentially decaying entanglement contours. We bring the reader's attention to \cite{2011PhRvB..84s5120G} where a similar quantity named the ``entropy density" is introduced to study the structure of entanglement entropy around the entangling surface in gapped phases. Similarly, the contour may be interesting to study in higher dimensional gapped phases.

\section{Higher dimensional contours, asymmetric subregions, and bit threads}

\label{bitthread}

\begin{figure}
    \centering
    \includegraphics[width = 5cm]{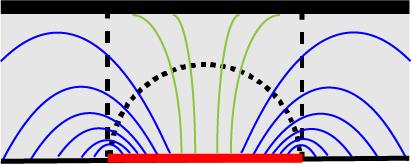}
    \caption{In the black hole geometry (subspace shown), there are two configurations for the minimal entanglement wedge cross section, the standard minimal surface (short dashed lines) and the disconnected surface reaching the horizon (longer dashes). For the entanglement entropy, bit threads can terminate on the black hole horizon, so both the green and blue bit threads contribute. For LN/EoP, only the blue bit threads will contribute.}
    \label{bip_LN}
\end{figure}

It is clearly of  interest to generalize the notion of an entanglement contour to arbitrary dimensions and entangling surface geometries. We address these issues in this section with motivations from the construction of holographic entanglement entropy from bit threads \cite{2017CMaPh.352..407F}. 
A bit thread construction involves a divergenceless vector field, $v$, whose norm is bounded by $(4 G_N)^{-1}$. We sketch a bit thread configuration in Fig. \ref{bip_LN}. The Ryu-Takayanagi formula for a boundary region, $A$, can then be reformulated as a maximization problem
\begin{align}
    S(A) = \max_v \int_A v,
\end{align}
where the maximization is over all bit thread configurations. By comparing to the normalization condition, we immediately see how this notion of bit threads is useful for constructing entanglement contours. For any maximal bit thread configuration respecting the spatial symmetries of the boundary region, we can define the contour function
\begin{align}
    s_A(x) = |v(x)|.
\end{align}

We now prove this obeys the conditions from \cite{2014JSMTE..10..011C}:
\begin{enumerate}
    \item Positivity: The contour is non-negative because the norm is non-negative.
    \item Normalization: The work of \cite{2017CMaPh.352..407F} explicitly proved the normalization of the contour over a boundary region using the max-flow min-cut theorem.
    \item Symmetry: We have imposed symmetry in the definition by only allowing bit thread configurations which respect the spatial symmetries of the region.
    \item Invariance under local unitaries: As before, this is ensured by causality.
    \item Upper bound: If the interval is a union of disconnected regions, there exist nontrivial factorizations of the Hilbert space. We take a disconnected region $A$ and a subregion $X \subset A$. The flux of bit threads at $X$ is bounded by the maximum flux of bit threads for all of $A$ over all bit thread configurations, which is equivalent to the von Neumann entropy of $A$. Therefore, the upper bound is satisfied.
\end{enumerate}

The degeneracy of extremal bit thread configurations is enormous. Given that these configurations respect the spatial symmetry of the boundary subregion, each provides a distinct contour function. This is a clear way of understanding the non-uniqueness of the entanglement contour. The only universal parts of the contour are the divergent pieces near the entangling surface. Far from the entangling surface, one can always choose a bit thread configuration for which the local contour function is trivial. However, the bit threads can be rearranged so that the local contour function is finite at the same point at which it was trivial for the other configuration. This is an unsettling phenomenon for a fine-grained entropy. We are willing to accept that the entanglement contour is degenerate, but wish to impose additional constraints that would enforce the following property for any two contours, $s^1_{A}$ and $s^2_{A}$: If $s^1_{A}(x) < s^1_{A}(y)$ for $x,y \in A$, then $s^2_{A}(x) < s^2_{A}(y)$. 

Happily, recent work on bit threads \cite{2018arXiv181108879A} has provided explicit constructions of bit thread configurations which satisfy our conditions for the contour. We begin with what the authors refer to as ``geodesic flows," where the bit threads follow bulk geodesics.

In $d$ spatial dimensions, for boundary regions with spherical entangling surfaces of radius $R$, the following norm of the geodesic flow was found
\begin{align}
    |v| = \frac{c}{6} \left(\frac{2 R z }{\sqrt{(R^2 + r^2 + z^2) - 4R^2 r^2}} \right)^d,
\end{align}
where $z$ is the bulk radial coordinate and $r$ is the distance from the center of the ball-shaped region on the boundary. Note that all angular dependence has dropped out. The contour is defined at the boundary, so we find
\begin{align}
    s_{A} (r) = \frac{c}{6} \left(\frac{2R}{R^2 - r^2} \right)^d.
\end{align}
% We plot this for several spatial dimensions in Fig. \ref{geo_bits}.
The dimensional dependence clearly shows that only for the 1+1d case
are there
significant entanglement contributions away from the entangling surface, displaying the logarithmic violation to the area law. Interestingly, for 1+1d, this is the exact ground state contour function that is found using the bulk modular flow (\ref{contour_eq}). This means that the bit threads following geodesics have matched boundary points to points on the RT surface identically to the matching from the modular flow.

\subsection{Entanglement wedge cross sections}
We would now like to restrict the bit threads to the entanglement wedge, which is the bulk dual of the CFT density matrix $\rho_{AB}$ \cite{2014arXiv1412.8465J}. The entanglement wedge is a codimension-1 bulk region whose boundary is the union of the RT surface and $A\cup B$.
 Using the formalism for holographic logarithmic negativity (LN) from \cite{2018arXiv180800446K} where the negativity is related to the area of the minimal entanglement wedge cross section, we can concoct a contour for negativity \footnote{An interesting attempt at constructing a negativity contour can be
  found in \cite{deNobili_thesis}}.
Minimal entanglement wedge cross-sections for general entangling surface geometries have also been studied in connection to entanglement of purification (EoP) \cite{2017arXiv170809393T, 2018JHEP...04..132B, 2018JHEP...01..098N, 2018arXiv180502625U,2018JHEP...03..006B,2018PTEP.2018f3B03H,2018PhRvD..98b6010N,2018arXiv180500476B,2018arXiv180405855E,2018arXiv181205268C,2019PhRvL.122n1601T}, so our contour can be interpreted in that context as well.
Both LN and EoP are useful measures for mixed state entanglement.

We define a holographic entanglement contour for LN and EoP much in the same way as we did for the contour for the entanglement entropy. Again, we need a divergenceless vector field, $v$, whose norm is now bounded by $\mathcal{X}_d (4 G_N)^{-1}$ for LN and $(4 G_N)^{-1}$ for EoP, where
\begin{align}
  \mathcal{X}_d &= \frac{1}{2}x_d^{d-2}\left(1 + x_d^2\right) - 1, 
  \\ \nonumber
   x_d &= \frac{2}{d} \left(1 + \sqrt{1 - \frac{d}{2} + \frac{d^2}{4}}\right), 
\end{align}
for spherical entangling surfaces. For less symmetric set-ups of LN, nontrivial backreactions in the bulk must be accounted for \cite{2018arXiv180800446K}. However, the formula for EoP is still valid for asymmetric set-ups.
Guaranteed by the max-flow min-cut theorem, the entanglement wedge cross-section and hence the LN and EoP are calculated as
\begin{align}
    \mathcal{E}(A,B) = \max_v \int_A v =  \max_v \int_B v,
\end{align}
where each bit thread must start and end on $A$ and $B$.
This leads us to the LN/EoP contour
\begin{align}
    e_{A} (x) = |v(x)|, \quad e_{B} (x) = |v(x)|.
\end{align}
In order to construct valid bit thread constructions restricted to the entanglement wedge, we need to use the notion of ``maximally packed flows" \cite{2018arXiv181108879A}. For a single interval $(-l/2, l/2)$ at inverse temperature $\beta$, there are two configurations for the entanglement wedge cross-section. The cross-over from the standard extremal surface to the one connected to the horizon occurs around $l/\beta \sim 0.28$ (Fig. \ref{bip_LN}). In the connected regime, the LN/EoP contour is simply proportional to the thermal entropy contour (\ref{therm_cont}) 
\begin{align}
    e_A(x) =  \frac{c \pi}{{\xi \beta}} \left(\coth\left(\frac{\pi (\frac{l}{2}+x )}{\beta} \right)+\coth\left(\frac{\pi (\frac{l}{2}-x)}{\beta} \right) \right),
\end{align}
where $\xi = 4$ for LN and $\xi = 6$ for EoP.
More interestingly, when the cross-section is disconnected, thermal contributions are subtracted and we find that the LN/EoP contour is only nontrivial (and divergent) near the entangling surface
\begin{align}
    e_A(x) = 0, \quad \frac{l}{2} - |x| > \frac{\beta}{4 \pi} \log 5
\end{align}
clearly showing that $\beta$ plays the role of the quantum correlation length in thermal systems.

\section{Connection to kinematic space}
\label{kinematic}

\begin{figure}
    \centering
    \includegraphics[width = 8cm]{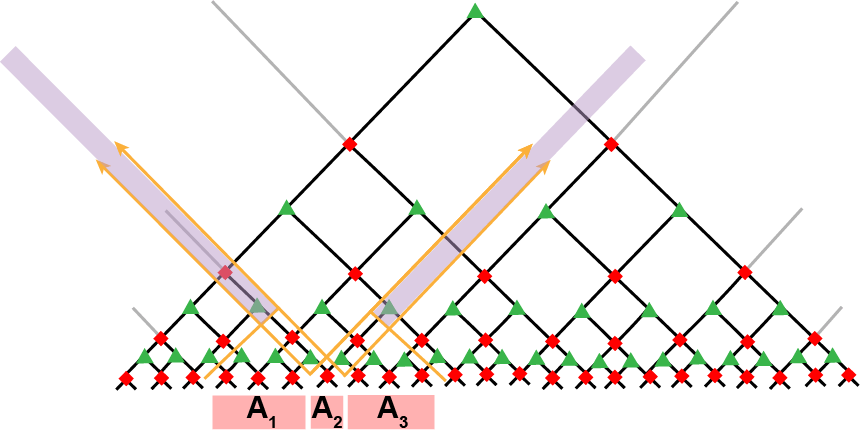}
    \caption{In kinematic space, the conditional mutual information is computed
      by the ``bulk" volumes. The left and right shaded lavender regions are
      $I(A_2, \bar{A} | A_1)$ and $I(A_2, \bar{A} | A_3)$ respectively.
      Interpreting MERA as kinematic space, the entanglement contour is computed
      by the number of isometries (green triangles) in the shaded lavender regions.
    %   \textcolor{red}{(Where are the shaded diamonds?)}
    }
    \label{kinMERA}
\end{figure}

Starting from a time slice of ${\it AdS}$ (or any other Euclidean surface), one
can define an auxiliary Lorentzian manifold known as ${\it kinematic \: space}$
\cite{2015JHEP...10..175C}. This manifold corresponds to the space of geodesics
on the original space.
In ${\it AdS}/$CFT, kinematic space encodes a number of important quantum
information theoretic quantities related to the boundary CFT state in an elegant
geometric fashion \cite{2015JHEP...10..175C}. Here we describe how the
entanglement contour can be encoded. Computationally, this may be useful for
investigating the local entanglement structure in MERA tensor networks, which
have been interpreted as a discretized version of kinematic space
\cite{2016JHEP...07..100C}. It also may be a useful local quantity for
reconstructing bulk geometry in ${\it AdS}/$CFT.

The length of a geodesic, $\gamma$ is computed by an integral over kinematic space, $K$
\begin{align}
    \gamma = \int_K \omega(u,v) m_{\gamma}(u,v),
\end{align}
where $\omega(u,v)$ is the Crofton form and $m_{\gamma}(u,v)$ is the number of intersections between $\gamma$ and the geodesic anchored on lightcone coordinates $(u,v)$.
Another quasilocal measure of entanglement called the \textit{entanglement density} is defined as \cite{2013JHEP...05..080N}
\begin{align}
    n(u,v) \equiv \frac{1}{2} \frac{\partial^2 S(u,v)}{\partial u \partial v}.
\end{align}
This serves as the natural form for kinematic space \cite{2015JHEP...10..175C}
\begin{equation}
    \omega(u,v) = \frac{\partial ^2 S(u,v)}{\partial u \partial v} du \wedge dv.
    \label{kinmet}
\end{equation}
Restricting ourselves to an overall pure state with periodic spatial boundary conditions, the entanglement contour is then an appropriate volume of kinematic space
\begin{align}
    s_A(x)= \frac{1}{2}\int_{\substack{\{(u,v) = (u,x)|u \in (x',x_1) \} \\ \cup \{(u,v) = (x,v)| v \in (x_2,x')\} }}  \omega,
\end{align}
with $x'$ unconstrained because the continuum version of (\ref{wen}) can be found by integrating the entanglement density
\begin{align}
    \int^{x_1}_{x'} n(x,y)dx+ \int_{x_2}^{x'} n(x,y)dy   \\ =\nonumber  \frac{1}{2}\left(\frac{\partial S(x_1, x)}{\partial x}-\frac{\partial S(x, x_2)}{\partial x} \right).
\end{align}
To choose a natural $x'$, we rewrite the entanglement contour in terms of conditional mutual information 
\begin{align}
    s_A(A_2) = \frac{1}{2}\left[ I(A_2, \bar{A} | A_{1})+I(A_2, \bar{A}| A_3 )\right],
\end{align}
agreeing with our initial construction of (\ref{contour_eq}).
Conditional mutual information is encoded in kinematic space by volumes of appropriate causal diamonds (multiplied by factors of $\log \chi$, where $\chi$ is the bond dimension). In MERA tensor networks, volumes are equivalent to the number of isometries in the region. We demonstrate this in Fig. \ref{kinMERA}.

The relation between the entanglement contour and conditional mutual information completes the circle of connections between the contour, entanglement density, bit threads, and kinematic space. It may be illuminating to study these connections further in the context of quantum bit threads in MERA \cite{ 2018arXiv180400441C}. 
\section{Quantum quenches}
\label{quench}

\begin{figure}
    \centering
    \includegraphics[width = 5cm]{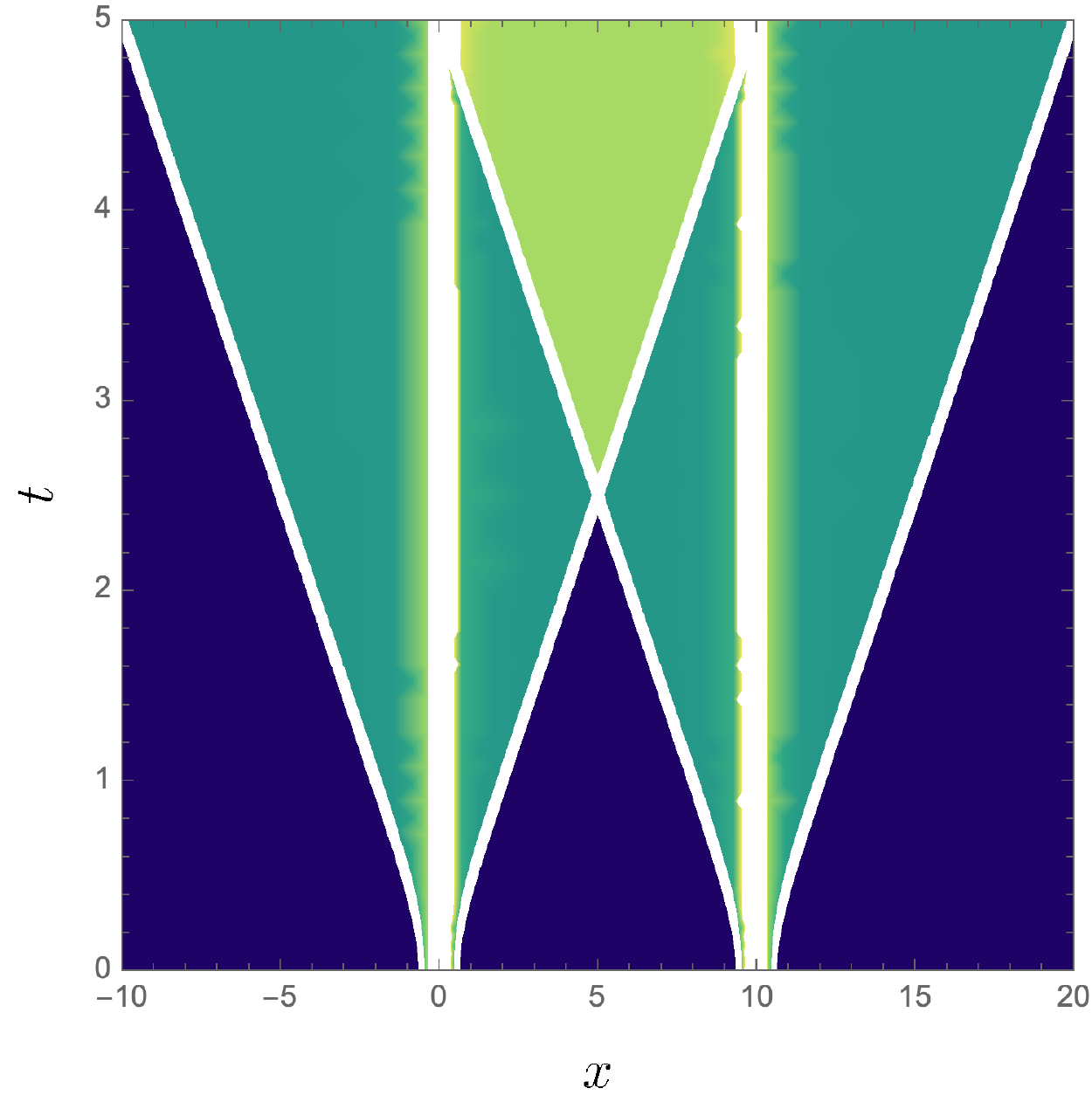}
    \caption{The entanglement contour following a global quench with $c = 1$ and $\beta = 2$. After initial quadratic growth, the contour waves propagate at $v_c = 2$ and cross one another at $t = l/4$, only to halt at $t = l/2$. The contour for the interval saturates at its thermal value.
    }
    \label{global_contour}
\end{figure}

The entanglement contour is particularly suited to dynamical settings because it can locally quantify how quantum information flows in time. We use quantum quenches to model out-of-equilibrium processes. Using (\ref{contour_eq}) for the contour,
% We denote quenches prepared by conformal maps of Riemann surfaces as \textit{Riemann quenches}. This distinguishes them from operator quenches, which we will discuss later.
we follow the formalism developed in \cite{2013arXiv1311.2562U} in which the Riemann surface corresponding to a given quantum quench is mapped to the half plane. From there, the holographic entanglement entropy can be computed by geodesics in AdS with a spacetime boundary.
For global and local (Calabrese-Cardy) quenches \cite{2006PhRvL..96m6801C,2007JSMTE..10....4C}, these conformal maps are
\begin{align}
    w_{glob}^{\pm}(x^{\pm}) &= e^{2 \pi x^{\pm}/\beta},
    \label{global_map} \\
    % w_{inhom}^{\pm}(x^{\pm}) &= \left( \frac{\sqrt{1 + 4 e^{2x^{\pm}/\lambda}} - 1}{2} \right)^{\pi \lambda/\beta}, \label{inhom_map}\\
        w_{loc}^{\pm}(x^{\pm}) &= \frac{x^{\pm}}{\epsilon} + \sqrt{\left(\frac{x^{\pm}}{\epsilon} \right)^2 + 1},
\label{loc_map}
\end{align}
where $\epsilon$ is a regulator. The contour for global quenches provides an excellent visual representation of the entanglement tsunami proposed in \cite{2014PhRvL.112a1601L}. This includes an early time quadratic growth and late time linear growth (see Fig. \ref{global_contour}). However, there is an important distinction between our picture and the tsunami. The tsunami velocity, $v_E$, for 1+1d CFTs is 1 in units of the speed of light. Similar to the discussion in Ref. \cite{2014JSMTE..10..011C}, we find that the entanglement propagates in real space at a \textit{contour velocity} of
\begin{align}
    v_c  = 2.
    \label{vc}
\end{align}
This is not a violation of causality because entanglement should not be thought of as a local object propagating from the entangling surface, as in the original tsunami picture. Rather, it can heuristically be thought of as a set of non-local waves generated at every point in space, analogous to the quasi-particle picture \cite{2005JSMTE..04..010C}. This leads to the larger ``velocity" of entanglement (\ref{vc}). Furthermore, the contour waves move through each other at $t = l/4$ and only halt once they have reached the opposite entangling surface at $t = l/2$. On the contrary, in the original tsunami picture, the waves stop once they have reached each other. This presents a significant difference when investigating the spatial structure of entanglement. The tsunami predicts that there is no entanglement between the center of the interval and the interval's complement when $t<l/2$ while the contour predicts nontrivial entanglement at the interval center for $t>l/4$. While the tsunami velocity accurately predicts the growth of the entanglement entropy non-locally (i.e. of the entire subsystem), the contour velocity captures the real-space velocity at which the entanglement spreads in real space. The contour velocity will generally depend on spacetime dimensions and its relation to the tsunami velocity will be less trivial than the factor of $2$ that was found in $1+1$d. It would be fascinating to find effective equations of motion for the entanglement contour analogous to those which have been studied for entanglement entropy and out-of-time-ordered correlators \cite{2013PhRvD..88b6012N,2017PhRvX...7c1016N,2018PhRvX...8b1014N,2018PhRvX...8b1013V,2018arXiv180300089J}. These may directly address the question of how quantum information locally flows in time. We expect different equations of motion for integrable and ergodic theories. 

\begin{figure}
    \centering
    \includegraphics[width = 4cm]{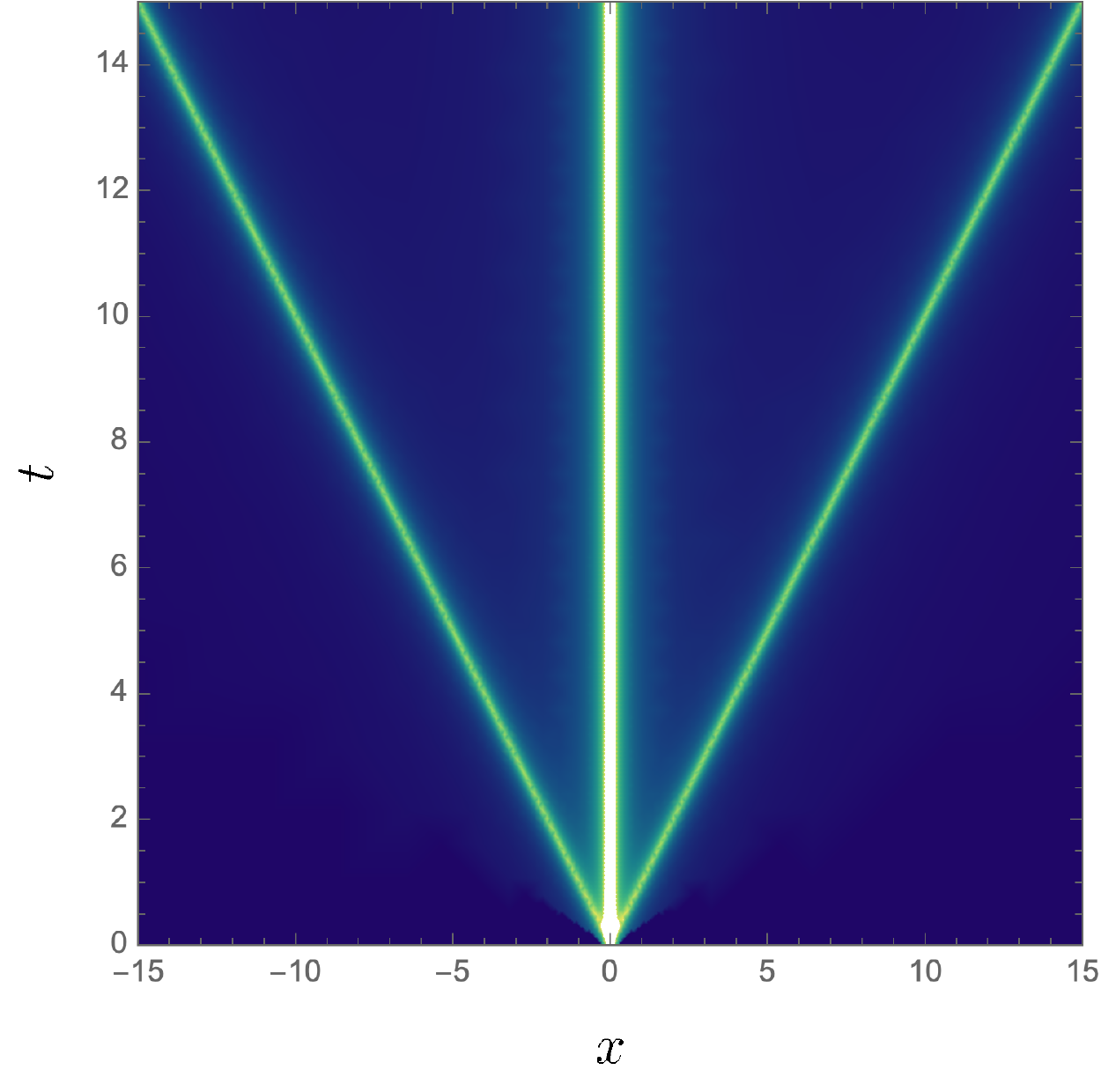}
    \includegraphics[width = 4cm]{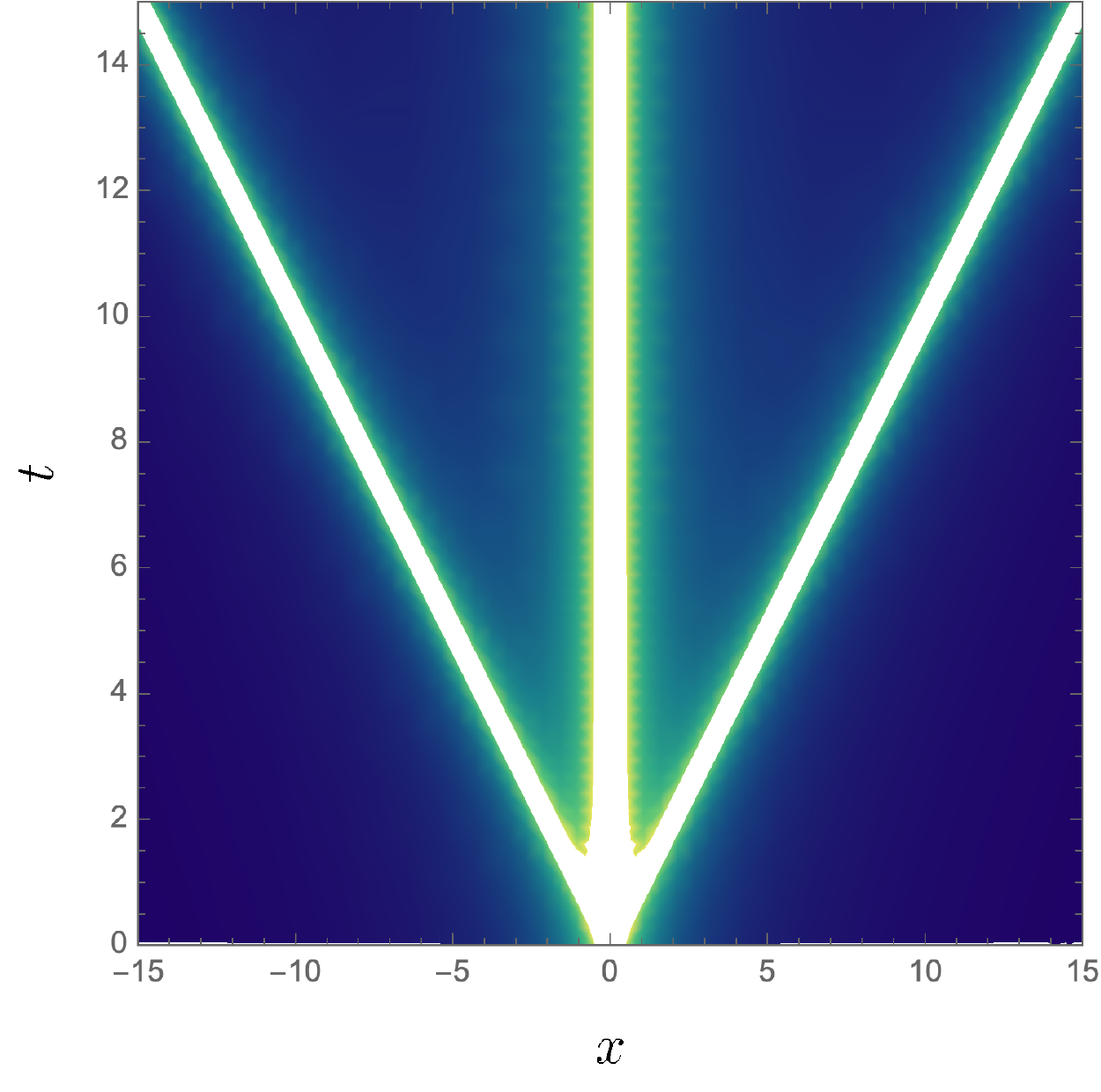}
    \caption{(left) The entanglement contour following a local (Calabrese-Cardy) quench for semi-infinite intervals with $c = 1$ and $\epsilon = 1/10$. (right) Local heavy operator quench with central charge $c = 1$, $\alpha_{\psi} = 1/2$, and $\delta = 1$. Now, $v_c = 1$. Once the wave front passes, the contour relaxes to its ground state value.}
    \label{loc_contour}
\end{figure}

For the local (Calabrese-Cardy) quench, the tsunami picture does not apply, though we find a transient contour wave propagating at $v_c = 1$.  (see Fig. \ref{loc_contour}). After the wave passes, the contour relaxes to its ground state value.

An additional way to induce a local quench is by inserting a local heavy operator of weight $h_{\psi}$ at the origin. The gravity dual to this protocol was shown to be a boosted black hole \cite{2013JHEP...05..080N}. 
The entanglement entropy for an interval $(l_1, l_2)$ away from the origin for a time $t>0$ is found \cite{2015JHEP...02..171A}
\begin{align}
  S^{out}_A= \begin{cases}
    \displaystyle
    \frac{c}{3} \log \left[ \frac{l_2-l_1}{\epsilon}\right]  &\hspace{-1.5cm} t<l_1 \mbox{ or } t>l_2\\
    \\
    \displaystyle
    \frac{c}{6}\log \left[ \frac{(l_2-l_1)(t-l_1)(l_2-t)}{\epsilon^2 \delta} \frac{\sin (\pi \alpha_{\psi})}{\alpha_{\psi}}\right] \\
     & \hspace{-1.2cm}l_1 <t<l_2
    \end{cases},
\end{align}
where $\delta$ is a regulator for the operator insertion and $\alpha_{\psi} = i \beta_{\psi}/{2 \pi}$. We find  an entanglement contour for semi-infinite intervals
\begin{align}
    s_A(x) = \begin{cases}
    \displaystyle
    \frac{c}{12}\frac{(2x-t)}{x(x-t)} & 0 < t<x \\ \\
    \displaystyle
    \frac{c}{6}\log\left[ \frac{\sin(\pi \alpha_{\psi})}{\delta \alpha_{\psi}}\right] \delta(x-t) & t = x \\ \\
    \displaystyle
    \frac{c}{6x} + \frac{c}{12}\frac{1}{t-x} & x < t
    \end{cases}.
\end{align}
The late time behavior is a combination of the vacuum value and a relaxation term. The contour following the operator quench is contrasted with the (Calabrese-Cardy) quench in Fig. \ref{loc_contour}.

For an interval that contains the origin, the entanglement entropy is

\begin{align}
  S^{in}_A= \begin{cases}
    \displaystyle
    \frac{c}{3} \log \left[ \frac{l_2-l_1}{\epsilon}\right] &\hspace{-1.3cm} t<|l_1|, t>l_2\\
    \\
    \displaystyle
    \frac{c}{6}\log \left[ \frac{(l_2-l_1)(l_1+t)(l_2+t)}{\epsilon^2 \delta} \frac{\sin (\pi \alpha_{\psi})}{\alpha_{\psi}}\right]
     \\& \hspace{-1.9cm}|l_1| <t<\sqrt{-l_1 l_2} \\ 
    \\
    \displaystyle
    \frac{c}{6}\log \left[ \frac{(l_2-l_1)(t-l_1)(l_2-t)}{\epsilon^2 \delta} \frac{\sin (\pi \alpha_{\psi})}{\alpha_{\psi}}\right]
    \\ &\hspace{-1.7cm} \sqrt{-l_1 l_2} <t<l_2
    \end{cases}.
\end{align}

An interesting feature of the contour can be seen from an interval containing the origin. As we would expect from causality, the total entanglement entropy of the interval is constant at early times. However, the contour is able to detect the rearrangement of entangled degrees of freedom within the interval. That is, the distribution of entanglement between the local degrees of freedom inside of the interval and the complement of the interval changes, but the total entanglement entropy of the interval remains the same.

\section{Discussion}
In this paper, we have introduced an entanglement contour for generic 1+1d systems and proposed a generalization to arbitrary dimensions and entangling surface geometries for holographic CFTs using the notion of bit threads. We emphasize that the entanglement contour, as specified by the five requirements, is generically a non-unique quantity, and our proposed form of the contour selects one particular contour function out of a possibly infinite number. States whose entanglement structure is purely bipartite (e.g. a completely dimerized state) are the only ones with a unique entanglement contour. We have found that the entanglement contour is particularly enlightening when studying dynamical situations and clarifies the physical picture associated to the entanglement tsunami. We argue that the tsunami velocity is the velocity of nonlocal entanglement growth for an interval while the contour velocity captures the real-space velocity of entanglement propagation in real space. It will be fascinating to study higher dimensional quantum quenches to determine how the contour velocity is affected. A covariant description of bit threads will play a key role. Furthermore, generalizations to higher curvature gravity (\`a la \cite{2018arXiv180704294H}), gauge theories, tensor network states other than MERA, as well as explicit constructions of negativity and purification contours are exciting avenues for understanding quantum entanglement at an even finer-grained level. Hopefully, this will lead to further insights about the connection between geometry and quantum entanglement. 
\acknowledgements{We thank Venkatesa Chandrasekaran, Paolo Glorioso, Hassan Shapourian, and Erik Tonni for useful discussion and comments. SR is supported by a Simons Investigator Grant from the Simons Foundation.}


\begin{thebibliography}{61}%
\makeatletter
\providecommand \@ifxundefined [1]{%
 \@ifx{#1\undefined}
}%
\providecommand \@ifnum [1]{%
 \ifnum #1\expandafter \@firstoftwo
 \else \expandafter \@secondoftwo
 \fi
}%
\providecommand \@ifx [1]{%
 \ifx #1\expandafter \@firstoftwo
 \else \expandafter \@secondoftwo
 \fi
}%
\providecommand \natexlab [1]{#1}%
\providecommand \enquote  [1]{``#1''}%
\providecommand \bibnamefont  [1]{#1}%
\providecommand \bibfnamefont [1]{#1}%
\providecommand \citenamefont [1]{#1}%
\providecommand \href@noop [0]{\@secondoftwo}%
\providecommand \href [0]{\begingroup \@sanitize@url \@href}%
\providecommand \@href[1]{\@@startlink{#1}\@@href}%
\providecommand \@@href[1]{\endgroup#1\@@endlink}%
\providecommand \@sanitize@url [0]{\catcode `\\12\catcode `\$12\catcode
  `\&12\catcode `\#12\catcode `\^12\catcode `\_12\catcode `\%12\relax}%
\providecommand \@@startlink[1]{}%
\providecommand \@@endlink[0]{}%
\providecommand \url  [0]{\begingroup\@sanitize@url \@url }%
\providecommand \@url [1]{\endgroup\@href {#1}{\urlprefix }}%
\providecommand \urlprefix  [0]{URL }%
\providecommand \Eprint [0]{\href }%
\providecommand \doibase [0]{http://dx.doi.org/}%
\providecommand \selectlanguage [0]{\@gobble}%
\providecommand \bibinfo  [0]{\@secondoftwo}%
\providecommand \bibfield  [0]{\@secondoftwo}%
\providecommand \translation [1]{[#1]}%
\providecommand \BibitemOpen [0]{}%
\providecommand \bibitemStop [0]{}%
\providecommand \bibitemNoStop [0]{.\EOS\space}%
\providecommand \EOS [0]{\spacefactor3000\relax}%
\providecommand \BibitemShut  [1]{\csname bibitem#1\endcsname}%
\let\auto@bib@innerbib\@empty
%</preamble>
\bibitem [{\citenamefont {{Vidal}}\ \emph {et~al.}(2003)\citenamefont
  {{Vidal}}, \citenamefont {{Latorre}}, \citenamefont {{Rico}},\ and\
  \citenamefont {{Kitaev}}}]{2003PhRvL..90v7902V}%
  \BibitemOpen
  \bibfield  {author} {\bibinfo {author} {\bibfnamefont {G.}~\bibnamefont
  {{Vidal}}}, \bibinfo {author} {\bibfnamefont {J.~I.}\ \bibnamefont
  {{Latorre}}}, \bibinfo {author} {\bibfnamefont {E.}~\bibnamefont {{Rico}}}, \
  and\ \bibinfo {author} {\bibfnamefont {A.}~\bibnamefont {{Kitaev}}},\ }\href
  {\doibase 10.1103/PhysRevLett.90.227902} {\bibfield  {journal} {\bibinfo
  {journal} {\prl}\ }\textbf {\bibinfo {volume} {90}},\ \bibinfo {eid} {227902}
  (\bibinfo {year} {2003})},\ \Eprint {http://arxiv.org/abs/quant-ph/0211074}
  {arXiv:quant-ph/0211074 [quant-ph]} \BibitemShut {NoStop}%
\bibitem [{\citenamefont {{Levin}}\ and\ \citenamefont
  {{Wen}}(2006)}]{2006PhRvL..96k0405L}%
  \BibitemOpen
  \bibfield  {author} {\bibinfo {author} {\bibfnamefont {M.}~\bibnamefont
  {{Levin}}}\ and\ \bibinfo {author} {\bibfnamefont {X.-G.}\ \bibnamefont
  {{Wen}}},\ }\href {\doibase 10.1103/PhysRevLett.96.110405} {\bibfield
  {journal} {\bibinfo  {journal} {\prl}\ }\textbf {\bibinfo {volume} {96}},\
  \bibinfo {eid} {110405} (\bibinfo {year} {2006})},\ \Eprint
  {http://arxiv.org/abs/cond-mat/0510613} {arXiv:cond-mat/0510613
  [cond-mat.str-el]} \BibitemShut {NoStop}%
\bibitem [{\citenamefont {Kitaev}\ and\ \citenamefont
  {Preskill}(2006)}]{PhysRevLett.96.110404}%
  \BibitemOpen
  \bibfield  {author} {\bibinfo {author} {\bibfnamefont {A.}~\bibnamefont
  {Kitaev}}\ and\ \bibinfo {author} {\bibfnamefont {J.}~\bibnamefont
  {Preskill}},\ }\href {\doibase 10.1103/PhysRevLett.96.110404} {\bibfield
  {journal} {\bibinfo  {journal} {Phys. Rev. Lett.}\ }\textbf {\bibinfo
  {volume} {96}},\ \bibinfo {pages} {110404} (\bibinfo {year}
  {2006})}\BibitemShut {NoStop}%
\bibitem [{\citenamefont {{Ryu}}\ and\ \citenamefont
  {{Takayanagi}}(2006{\natexlab{a}})}]{2006PhRvL..96r1602R}%
  \BibitemOpen
  \bibfield  {author} {\bibinfo {author} {\bibfnamefont {S.}~\bibnamefont
  {{Ryu}}}\ and\ \bibinfo {author} {\bibfnamefont {T.}~\bibnamefont
  {{Takayanagi}}},\ }\href {\doibase 10.1103/PhysRevLett.96.181602} {\bibfield
  {journal} {\bibinfo  {journal} {Physical Review Letters}\ }\textbf {\bibinfo
  {volume} {96}},\ \bibinfo {eid} {181602} (\bibinfo {year}
  {2006}{\natexlab{a}})},\ \Eprint {http://arxiv.org/abs/hep-th/0603001}
  {hep-th/0603001} \BibitemShut {NoStop}%
\bibitem [{\citenamefont {{Ryu}}\ and\ \citenamefont
  {{Takayanagi}}(2006{\natexlab{b}})}]{2006JHEP...08..045R}%
  \BibitemOpen
  \bibfield  {author} {\bibinfo {author} {\bibfnamefont {S.}~\bibnamefont
  {{Ryu}}}\ and\ \bibinfo {author} {\bibfnamefont {T.}~\bibnamefont
  {{Takayanagi}}},\ }\href {\doibase 10.1088/1126-6708/2006/08/045} {\bibfield
  {journal} {\bibinfo  {journal} {Journal of High Energy Physics}\ }\textbf
  {\bibinfo {volume} {8}},\ \bibinfo {eid} {045} (\bibinfo {year}
  {2006}{\natexlab{b}})},\ \Eprint {http://arxiv.org/abs/hep-th/0605073}
  {hep-th/0605073} \BibitemShut {NoStop}%
\bibitem [{\citenamefont {{van Raamsdonk}}(2010)}]{2010GReGr..42.2323V}%
  \BibitemOpen
  \bibfield  {author} {\bibinfo {author} {\bibfnamefont {M.}~\bibnamefont {{van
  Raamsdonk}}},\ }\href {\doibase 10.1007/s10714-010-1034-0} {\bibfield
  {journal} {\bibinfo  {journal} {General Relativity and Gravitation}\ }\textbf
  {\bibinfo {volume} {42}},\ \bibinfo {pages} {2323} (\bibinfo {year}
  {2010})},\ \Eprint {http://arxiv.org/abs/1005.3035} {arXiv:1005.3035
  [hep-th]} \BibitemShut {NoStop}%
\bibitem [{\citenamefont {{Hubeny}}\ \emph {et~al.}(2007)\citenamefont
  {{Hubeny}}, \citenamefont {{Rangamani}},\ and\ \citenamefont
  {{Takayanagi}}}]{2007JHEP...07..062H}%
  \BibitemOpen
  \bibfield  {author} {\bibinfo {author} {\bibfnamefont {V.~E.}\ \bibnamefont
  {{Hubeny}}}, \bibinfo {author} {\bibfnamefont {M.}~\bibnamefont
  {{Rangamani}}}, \ and\ \bibinfo {author} {\bibfnamefont {T.}~\bibnamefont
  {{Takayanagi}}},\ }\href {\doibase 10.1088/1126-6708/2007/07/062} {\bibfield
  {journal} {\bibinfo  {journal} {Journal of High Energy Physics}\ }\textbf
  {\bibinfo {volume} {2007}},\ \bibinfo {eid} {062} (\bibinfo {year} {2007})},\
  \Eprint {http://arxiv.org/abs/0705.0016} {arXiv:0705.0016 [hep-th]}
  \BibitemShut {NoStop}%
\bibitem [{\citenamefont {{Chen}}\ and\ \citenamefont
  {{Vidal}}(2014)}]{2014JSMTE..10..011C}%
  \BibitemOpen
  \bibfield  {author} {\bibinfo {author} {\bibfnamefont {Y.}~\bibnamefont
  {{Chen}}}\ and\ \bibinfo {author} {\bibfnamefont {G.}~\bibnamefont
  {{Vidal}}},\ }\href {\doibase 10.1088/1742-5468/2014/10/P10011} {\bibfield
  {journal} {\bibinfo  {journal} {Journal of Statistical Mechanics: Theory and
  Experiment}\ }\textbf {\bibinfo {volume} {10}},\ \bibinfo {eid} {10011}
  (\bibinfo {year} {2014})},\ \Eprint {http://arxiv.org/abs/1406.1471}
  {arXiv:1406.1471 [cond-mat.str-el]} \BibitemShut {NoStop}%
\bibitem [{Note1()}]{Note1}%
  \BibitemOpen
  \bibinfo {note} {We note that a similar construction was studied earlier in
  Ref. \cite {2004PhRvA..70e2329B}.}\BibitemShut {Stop}%
\bibitem [{\citenamefont {{Coser}}\ \emph {et~al.}(2017)\citenamefont
  {{Coser}}, \citenamefont {{De Nobili}},\ and\ \citenamefont
  {{Tonni}}}]{2017JPhA...50E4001C}%
  \BibitemOpen
  \bibfield  {author} {\bibinfo {author} {\bibfnamefont {A.}~\bibnamefont
  {{Coser}}}, \bibinfo {author} {\bibfnamefont {C.}~\bibnamefont {{De
  Nobili}}}, \ and\ \bibinfo {author} {\bibfnamefont {E.}~\bibnamefont
  {{Tonni}}},\ }\href {\doibase 10.1088/1751-8121/aa7902} {\bibfield  {journal}
  {\bibinfo  {journal} {Journal of Physics A Mathematical General}\ }\textbf
  {\bibinfo {volume} {50}},\ \bibinfo {eid} {314001} (\bibinfo {year}
  {2017})},\ \Eprint {http://arxiv.org/abs/1701.08427} {arXiv:1701.08427
  [cond-mat.stat-mech]} \BibitemShut {NoStop}%
\bibitem [{\citenamefont {{Wen}}(2018{\natexlab{a}})}]{2018arXiv180305552W}%
  \BibitemOpen
  \bibfield  {author} {\bibinfo {author} {\bibfnamefont {Q.}~\bibnamefont
  {{Wen}}},\ }\href@noop {} {\bibfield  {journal} {\bibinfo  {journal} {ArXiv
  e-prints}\ } (\bibinfo {year} {2018}{\natexlab{a}})},\ \Eprint
  {http://arxiv.org/abs/1803.05552} {arXiv:1803.05552 [hep-th]} \BibitemShut
  {NoStop}%
\bibitem [{\citenamefont {{Alba}}\ \emph {et~al.}(2018)\citenamefont {{Alba}},
  \citenamefont {{Santalla}}, \citenamefont {{Ruggiero}}, \citenamefont
  {{Rodriguez-Laguna}}, \citenamefont {{Calabrese}},\ and\ \citenamefont
  {{Sierra}}}]{2018arXiv180704179A}%
  \BibitemOpen
  \bibfield  {author} {\bibinfo {author} {\bibfnamefont {V.}~\bibnamefont
  {{Alba}}}, \bibinfo {author} {\bibfnamefont {S.~N.}\ \bibnamefont
  {{Santalla}}}, \bibinfo {author} {\bibfnamefont {P.}~\bibnamefont
  {{Ruggiero}}}, \bibinfo {author} {\bibfnamefont {J.}~\bibnamefont
  {{Rodriguez-Laguna}}}, \bibinfo {author} {\bibfnamefont {P.}~\bibnamefont
  {{Calabrese}}}, \ and\ \bibinfo {author} {\bibfnamefont {G.}~\bibnamefont
  {{Sierra}}},\ }\href@noop {} {\bibfield  {journal} {\bibinfo  {journal}
  {ArXiv e-prints}\ } (\bibinfo {year} {2018})},\ \Eprint
  {http://arxiv.org/abs/1807.04179} {arXiv:1807.04179 [cond-mat.str-el]}
  \BibitemShut {NoStop}%
\bibitem [{\citenamefont {{Tonni}}\ \emph {et~al.}(2018)\citenamefont
  {{Tonni}}, \citenamefont {{Rodr{\'{\i}}guez-Laguna}},\ and\ \citenamefont
  {{Sierra}}}]{2018JSMTE..04.3105T}%
  \BibitemOpen
  \bibfield  {author} {\bibinfo {author} {\bibfnamefont {E.}~\bibnamefont
  {{Tonni}}}, \bibinfo {author} {\bibfnamefont {J.}~\bibnamefont
  {{Rodr{\'{\i}}guez-Laguna}}}, \ and\ \bibinfo {author} {\bibfnamefont
  {G.}~\bibnamefont {{Sierra}}},\ }\href {\doibase 10.1088/1742-5468/aab67d}
  {\bibfield  {journal} {\bibinfo  {journal} {Journal of Statistical Mechanics:
  Theory and Experiment}\ }\textbf {\bibinfo {volume} {4}},\ \bibinfo {pages}
  {043105} (\bibinfo {year} {2018})},\ \Eprint
  {http://arxiv.org/abs/1712.03557} {arXiv:1712.03557 [cond-mat.stat-mech]}
  \BibitemShut {NoStop}%
\bibitem [{\citenamefont {{Freedman}}\ and\ \citenamefont
  {{Headrick}}(2017)}]{2017CMaPh.352..407F}%
  \BibitemOpen
  \bibfield  {author} {\bibinfo {author} {\bibfnamefont {M.}~\bibnamefont
  {{Freedman}}}\ and\ \bibinfo {author} {\bibfnamefont {M.}~\bibnamefont
  {{Headrick}}},\ }\href {\doibase 10.1007/s00220-016-2796-3} {\bibfield
  {journal} {\bibinfo  {journal} {Communications in Mathematical Physics}\
  }\textbf {\bibinfo {volume} {352}},\ \bibinfo {pages} {407} (\bibinfo {year}
  {2017})},\ \Eprint {http://arxiv.org/abs/1604.00354} {arXiv:1604.00354
  [hep-th]} \BibitemShut {NoStop}%
\bibitem [{Note2()}]{Note2}%
  \BibitemOpen
  \bibinfo {note} {We also thank Erik Tonni for directing us to his slides from
  It from Qubit Bariloche where a connection between bit threads and the
  entanglement contour was originally discussed.}\BibitemShut {Stop}%
\bibitem [{\citenamefont {{Liu}}\ and\ \citenamefont
  {{Suh}}(2014{\natexlab{a}})}]{2014PhRvL.112a1601L}%
  \BibitemOpen
  \bibfield  {author} {\bibinfo {author} {\bibfnamefont {H.}~\bibnamefont
  {{Liu}}}\ and\ \bibinfo {author} {\bibfnamefont {S.~J.}\ \bibnamefont
  {{Suh}}},\ }\href {\doibase 10.1103/PhysRevLett.112.011601} {\bibfield
  {journal} {\bibinfo  {journal} {Physical Review Letters}\ }\textbf {\bibinfo
  {volume} {112}},\ \bibinfo {eid} {011601} (\bibinfo {year}
  {2014}{\natexlab{a}})},\ \Eprint {http://arxiv.org/abs/1305.7244}
  {arXiv:1305.7244 [hep-th]} \BibitemShut {NoStop}%
\bibitem [{\citenamefont {{Liu}}\ and\ \citenamefont
  {{Suh}}(2014{\natexlab{b}})}]{2014PhRvD..89f6012L}%
  \BibitemOpen
  \bibfield  {author} {\bibinfo {author} {\bibfnamefont {H.}~\bibnamefont
  {{Liu}}}\ and\ \bibinfo {author} {\bibfnamefont {S.~J.}\ \bibnamefont
  {{Suh}}},\ }\href {\doibase 10.1103/PhysRevD.89.066012} {\bibfield  {journal}
  {\bibinfo  {journal} {\prd}\ }\textbf {\bibinfo {volume} {89}},\ \bibinfo
  {eid} {066012} (\bibinfo {year} {2014}{\natexlab{b}})},\ \Eprint
  {http://arxiv.org/abs/1311.1200} {arXiv:1311.1200 [hep-th]} \BibitemShut
  {NoStop}%
\bibitem [{\citenamefont {{Casini}}\ \emph {et~al.}(2016)\citenamefont
  {{Casini}}, \citenamefont {{Liu}},\ and\ \citenamefont
  {{Mezei}}}]{2016JHEP...07..077C}%
  \BibitemOpen
  \bibfield  {author} {\bibinfo {author} {\bibfnamefont {H.}~\bibnamefont
  {{Casini}}}, \bibinfo {author} {\bibfnamefont {H.}~\bibnamefont {{Liu}}}, \
  and\ \bibinfo {author} {\bibfnamefont {M.}~\bibnamefont {{Mezei}}},\ }\href
  {\doibase 10.1007/JHEP07(2016)077} {\bibfield  {journal} {\bibinfo  {journal}
  {Journal of High Energy Physics}\ }\textbf {\bibinfo {volume} {2016}},\
  \bibinfo {eid} {77} (\bibinfo {year} {2016})},\ \Eprint
  {http://arxiv.org/abs/1509.05044} {arXiv:1509.05044 [hep-th]} \BibitemShut
  {NoStop}%
\bibitem [{\citenamefont {{Leichenauer}}\ and\ \citenamefont
  {{Moosa}}(2015)}]{2015PhRvD..92l6004L}%
  \BibitemOpen
  \bibfield  {author} {\bibinfo {author} {\bibfnamefont {S.}~\bibnamefont
  {{Leichenauer}}}\ and\ \bibinfo {author} {\bibfnamefont {M.}~\bibnamefont
  {{Moosa}}},\ }\href {\doibase 10.1103/PhysRevD.92.126004} {\bibfield
  {journal} {\bibinfo  {journal} {\prd}\ }\textbf {\bibinfo {volume} {92}},\
  \bibinfo {eid} {126004} (\bibinfo {year} {2015})},\ \Eprint
  {http://arxiv.org/abs/1505.04225} {arXiv:1505.04225 [hep-th]} \BibitemShut
  {NoStop}%
\bibitem [{\citenamefont {Araki}\ and\ \citenamefont {Lieb}(1970)}]{araki1970}%
  \BibitemOpen
  \bibfield  {author} {\bibinfo {author} {\bibfnamefont {H.}~\bibnamefont
  {Araki}}\ and\ \bibinfo {author} {\bibfnamefont {E.~H.}\ \bibnamefont
  {Lieb}},\ }\href {https://projecteuclid.org:443/euclid.cmp/1103842506}
  {\bibfield  {journal} {\bibinfo  {journal} {Comm. Math. Phys.}\ }\textbf
  {\bibinfo {volume} {18}},\ \bibinfo {pages} {160} (\bibinfo {year}
  {1970})}\BibitemShut {NoStop}%
\bibitem [{\citenamefont {{Wen}}(2018{\natexlab{b}})}]{2018arXiv181011756W}%
  \BibitemOpen
  \bibfield  {author} {\bibinfo {author} {\bibfnamefont {Q.}~\bibnamefont
  {{Wen}}},\ }\href@noop {} {\bibfield  {journal} {\bibinfo  {journal} {ArXiv
  e-prints}\ ,\ \bibinfo {eid} {arXiv:1810.11756}} (\bibinfo {year}
  {2018}{\natexlab{b}})},\ \Eprint {http://arxiv.org/abs/1810.11756}
  {arXiv:1810.11756 [hep-th]} \BibitemShut {NoStop}%
\bibitem [{\citenamefont {{Comp{\`e}re}}\ \emph {et~al.}(2013)\citenamefont
  {{Comp{\`e}re}}, \citenamefont {{Song}},\ and\ \citenamefont
  {{Strominger}}}]{2013JHEP...05..152C}%
  \BibitemOpen
  \bibfield  {author} {\bibinfo {author} {\bibfnamefont {G.}~\bibnamefont
  {{Comp{\`e}re}}}, \bibinfo {author} {\bibfnamefont {W.}~\bibnamefont
  {{Song}}}, \ and\ \bibinfo {author} {\bibfnamefont {A.}~\bibnamefont
  {{Strominger}}},\ }\href {\doibase 10.1007/JHEP05(2013)152} {\bibfield
  {journal} {\bibinfo  {journal} {Journal of High Energy Physics}\ }\textbf
  {\bibinfo {volume} {2013}},\ \bibinfo {eid} {152} (\bibinfo {year} {2013})},\
  \Eprint {http://arxiv.org/abs/1303.2662} {arXiv:1303.2662 [hep-th]}
  \BibitemShut {NoStop}%
\bibitem [{\citenamefont {{Takayanagi}}(2011)}]{2011PhRvL.107j1602T}%
  \BibitemOpen
  \bibfield  {author} {\bibinfo {author} {\bibfnamefont {T.}~\bibnamefont
  {{Takayanagi}}},\ }\href {\doibase 10.1103/PhysRevLett.107.101602} {\bibfield
   {journal} {\bibinfo  {journal} {Physical Review Letters}\ }\textbf {\bibinfo
  {volume} {107}},\ \bibinfo {eid} {101602} (\bibinfo {year} {2011})},\ \Eprint
  {http://arxiv.org/abs/1105.5165} {arXiv:1105.5165 [hep-th]} \BibitemShut
  {NoStop}%
\bibitem [{\citenamefont {{Azeyanagi}}\ \emph {et~al.}(2008)\citenamefont
  {{Azeyanagi}}, \citenamefont {{Takayanagi}}, \citenamefont {{Karch}},\ and\
  \citenamefont {{Thompson}}}]{2008JHEP...03..054A}%
  \BibitemOpen
  \bibfield  {author} {\bibinfo {author} {\bibfnamefont {T.}~\bibnamefont
  {{Azeyanagi}}}, \bibinfo {author} {\bibfnamefont {T.}~\bibnamefont
  {{Takayanagi}}}, \bibinfo {author} {\bibfnamefont {A.}~\bibnamefont
  {{Karch}}}, \ and\ \bibinfo {author} {\bibfnamefont {E.~G.}\ \bibnamefont
  {{Thompson}}},\ }\href {\doibase 10.1088/1126-6708/2008/03/054} {\bibfield
  {journal} {\bibinfo  {journal} {Journal of High Energy Physics}\ }\textbf
  {\bibinfo {volume} {3}},\ \bibinfo {eid} {054} (\bibinfo {year} {2008})},\
  \Eprint {http://arxiv.org/abs/0712.1850} {arXiv:0712.1850 [hep-th]}
  \BibitemShut {NoStop}%
\bibitem [{\citenamefont {{Fitzpatrick}}\ \emph {et~al.}(2014)\citenamefont
  {{Fitzpatrick}}, \citenamefont {{Kaplan}},\ and\ \citenamefont
  {{Walters}}}]{2014JHEP...08..145F}%
  \BibitemOpen
  \bibfield  {author} {\bibinfo {author} {\bibfnamefont {A.~L.}\ \bibnamefont
  {{Fitzpatrick}}}, \bibinfo {author} {\bibfnamefont {J.}~\bibnamefont
  {{Kaplan}}}, \ and\ \bibinfo {author} {\bibfnamefont {M.~T.}\ \bibnamefont
  {{Walters}}},\ }\href {\doibase 10.1007/JHEP08(2014)145} {\bibfield
  {journal} {\bibinfo  {journal} {Journal of High Energy Physics}\ }\textbf
  {\bibinfo {volume} {8}},\ \bibinfo {eid} {145} (\bibinfo {year} {2014})},\
  \Eprint {http://arxiv.org/abs/1403.6829} {arXiv:1403.6829 [hep-th]}
  \BibitemShut {NoStop}%
\bibitem [{\citenamefont {{Asplund}}\ \emph {et~al.}(2015)\citenamefont
  {{Asplund}}, \citenamefont {{Bernamonti}}, \citenamefont {{Galli}},\ and\
  \citenamefont {{Hartman}}}]{2015JHEP...02..171A}%
  \BibitemOpen
  \bibfield  {author} {\bibinfo {author} {\bibfnamefont {C.~T.}\ \bibnamefont
  {{Asplund}}}, \bibinfo {author} {\bibfnamefont {A.}~\bibnamefont
  {{Bernamonti}}}, \bibinfo {author} {\bibfnamefont {F.}~\bibnamefont
  {{Galli}}}, \ and\ \bibinfo {author} {\bibfnamefont {T.}~\bibnamefont
  {{Hartman}}},\ }\href {\doibase 10.1007/JHEP02(2015)171} {\bibfield
  {journal} {\bibinfo  {journal} {Journal of High Energy Physics}\ }\textbf
  {\bibinfo {volume} {2}},\ \bibinfo {eid} {171} (\bibinfo {year} {2015})},\
  \Eprint {http://arxiv.org/abs/1410.1392} {arXiv:1410.1392 [hep-th]}
  \BibitemShut {NoStop}%
\bibitem [{\citenamefont {{Caputa}}\ \emph {et~al.}(2015)\citenamefont
  {{Caputa}}, \citenamefont {{Sim{\'o}n}}, \citenamefont {{{\v{S}}tikonas}},\
  and\ \citenamefont {{Takayanagi}}}]{2015JHEP...01..102C}%
  \BibitemOpen
  \bibfield  {author} {\bibinfo {author} {\bibfnamefont {P.}~\bibnamefont
  {{Caputa}}}, \bibinfo {author} {\bibfnamefont {J.}~\bibnamefont
  {{Sim{\'o}n}}}, \bibinfo {author} {\bibfnamefont {A.}~\bibnamefont
  {{{\v{S}}tikonas}}}, \ and\ \bibinfo {author} {\bibfnamefont
  {T.}~\bibnamefont {{Takayanagi}}},\ }\href {\doibase 10.1007/JHEP01(2015)102}
  {\bibfield  {journal} {\bibinfo  {journal} {Journal of High Energy Physics}\
  }\textbf {\bibinfo {volume} {2015}},\ \bibinfo {eid} {102} (\bibinfo {year}
  {2015})},\ \Eprint {http://arxiv.org/abs/1410.2287} {arXiv:1410.2287
  [hep-th]} \BibitemShut {NoStop}%
\bibitem [{\citenamefont {Srednicki}(1994)}]{PhysRevE.50.888}%
  \BibitemOpen
  \bibfield  {author} {\bibinfo {author} {\bibfnamefont {M.}~\bibnamefont
  {Srednicki}},\ }\href {\doibase 10.1103/PhysRevE.50.888} {\bibfield
  {journal} {\bibinfo  {journal} {Phys. Rev. E}\ }\textbf {\bibinfo {volume}
  {50}},\ \bibinfo {pages} {888} (\bibinfo {year} {1994})}\BibitemShut
  {NoStop}%
\bibitem [{\citenamefont {{Turner}}\ \emph {et~al.}(2018)\citenamefont
  {{Turner}}, \citenamefont {{Michailidis}}, \citenamefont {{Abanin}},
  \citenamefont {{Serbyn}},\ and\ \citenamefont
  {{Papi{\'c}}}}]{2018NatPh..14..745T}%
  \BibitemOpen
  \bibfield  {author} {\bibinfo {author} {\bibfnamefont {C.~J.}\ \bibnamefont
  {{Turner}}}, \bibinfo {author} {\bibfnamefont {A.~A.}\ \bibnamefont
  {{Michailidis}}}, \bibinfo {author} {\bibfnamefont {D.~A.}\ \bibnamefont
  {{Abanin}}}, \bibinfo {author} {\bibfnamefont {M.}~\bibnamefont {{Serbyn}}},
  \ and\ \bibinfo {author} {\bibfnamefont {Z.}~\bibnamefont {{Papi{\'c}}}},\
  }\href {\doibase 10.1038/s41567-018-0137-5} {\bibfield  {journal} {\bibinfo
  {journal} {Nature Physics}\ }\textbf {\bibinfo {volume} {14}},\ \bibinfo
  {pages} {745} (\bibinfo {year} {2018})}\BibitemShut {NoStop}%
\bibitem [{\citenamefont {{Grover}}\ \emph {et~al.}(2011)\citenamefont
  {{Grover}}, \citenamefont {{Turner}},\ and\ \citenamefont
  {{Vishwanath}}}]{2011PhRvB..84s5120G}%
  \BibitemOpen
  \bibfield  {author} {\bibinfo {author} {\bibfnamefont {T.}~\bibnamefont
  {{Grover}}}, \bibinfo {author} {\bibfnamefont {A.~M.}\ \bibnamefont
  {{Turner}}}, \ and\ \bibinfo {author} {\bibfnamefont {A.}~\bibnamefont
  {{Vishwanath}}},\ }\href {\doibase 10.1103/PhysRevB.84.195120} {\bibfield
  {journal} {\bibinfo  {journal} {Physical Review B}\ }\textbf {\bibinfo
  {volume} {84}},\ \bibinfo {eid} {195120} (\bibinfo {year} {2011})},\ \Eprint
  {http://arxiv.org/abs/1108.4038} {arXiv:1108.4038 [cond-mat.str-el]}
  \BibitemShut {NoStop}%
\bibitem [{\citenamefont {{Ag{\'o}n}}\ \emph {et~al.}(2018)\citenamefont
  {{Ag{\'o}n}}, \citenamefont {{de Boer}},\ and\ \citenamefont
  {{Pedraza}}}]{2018arXiv181108879A}%
  \BibitemOpen
  \bibfield  {author} {\bibinfo {author} {\bibfnamefont {C.~A.}\ \bibnamefont
  {{Ag{\'o}n}}}, \bibinfo {author} {\bibfnamefont {J.}~\bibnamefont {{de
  Boer}}}, \ and\ \bibinfo {author} {\bibfnamefont {J.~F.}\ \bibnamefont
  {{Pedraza}}},\ }\href@noop {} {\bibfield  {journal} {\bibinfo  {journal}
  {ArXiv e-prints}\ } (\bibinfo {year} {2018})},\ \Eprint
  {http://arxiv.org/abs/1811.08879} {arXiv:1811.08879 [hep-th]} \BibitemShut
  {NoStop}%
\bibitem [{\citenamefont {{Jafferis}}\ and\ \citenamefont
  {{Suh}}(2014)}]{2014arXiv1412.8465J}%
  \BibitemOpen
  \bibfield  {author} {\bibinfo {author} {\bibfnamefont {D.~L.}\ \bibnamefont
  {{Jafferis}}}\ and\ \bibinfo {author} {\bibfnamefont {S.~J.}\ \bibnamefont
  {{Suh}}},\ }\href@noop {} {\bibfield  {journal} {\bibinfo  {journal} {ArXiv
  e-prints}\ ,\ \bibinfo {eid} {arXiv:1412.8465}} (\bibinfo {year} {2014})},\
  \Eprint {http://arxiv.org/abs/1412.8465} {arXiv:1412.8465 [hep-th]}
  \BibitemShut {NoStop}%
\bibitem [{\citenamefont {{Kudler-Flam}}\ and\ \citenamefont
  {{Ryu}}(2018)}]{2018arXiv180800446K}%
  \BibitemOpen
  \bibfield  {author} {\bibinfo {author} {\bibfnamefont {J.}~\bibnamefont
  {{Kudler-Flam}}}\ and\ \bibinfo {author} {\bibfnamefont {S.}~\bibnamefont
  {{Ryu}}},\ }\href@noop {} {\bibfield  {journal} {\bibinfo  {journal} {ArXiv
  e-prints}\ } (\bibinfo {year} {2018})},\ \Eprint
  {http://arxiv.org/abs/1808.00446} {arXiv:1808.00446 [hep-th]} \BibitemShut
  {NoStop}%
\bibitem [{Note3()}]{Note3}%
  \BibitemOpen
  \bibinfo {note} {An interesting attempt at constructing a negativity contour
  can be found in \cite {deNobili_thesis}}\BibitemShut {NoStop}%
\bibitem [{\citenamefont {{Takayanagi}}\ and\ \citenamefont
  {{Umemoto}}(2017)}]{2017arXiv170809393T}%
  \BibitemOpen
  \bibfield  {author} {\bibinfo {author} {\bibfnamefont {T.}~\bibnamefont
  {{Takayanagi}}}\ and\ \bibinfo {author} {\bibfnamefont {K.}~\bibnamefont
  {{Umemoto}}},\ }\href@noop {} {\bibfield  {journal} {\bibinfo  {journal}
  {ArXiv e-prints}\ } (\bibinfo {year} {2017})},\ \Eprint
  {http://arxiv.org/abs/1708.09393} {arXiv:1708.09393 [hep-th]} \BibitemShut
  {NoStop}%
\bibitem [{\citenamefont {{Bhattacharyya}}\ \emph {et~al.}(2018)\citenamefont
  {{Bhattacharyya}}, \citenamefont {{Takayanagi}},\ and\ \citenamefont
  {{Umemoto}}}]{2018JHEP...04..132B}%
  \BibitemOpen
  \bibfield  {author} {\bibinfo {author} {\bibfnamefont {A.}~\bibnamefont
  {{Bhattacharyya}}}, \bibinfo {author} {\bibfnamefont {T.}~\bibnamefont
  {{Takayanagi}}}, \ and\ \bibinfo {author} {\bibfnamefont {K.}~\bibnamefont
  {{Umemoto}}},\ }\href {\doibase 10.1007/JHEP04(2018)132} {\bibfield
  {journal} {\bibinfo  {journal} {Journal of High Energy Physics}\ }\textbf
  {\bibinfo {volume} {4}},\ \bibinfo {eid} {132} (\bibinfo {year} {2018})},\
  \Eprint {http://arxiv.org/abs/1802.09545} {arXiv:1802.09545 [hep-th]}
  \BibitemShut {NoStop}%
\bibitem [{\citenamefont {{Nguyen}}\ \emph {et~al.}(2018)\citenamefont
  {{Nguyen}}, \citenamefont {{Devakul}}, \citenamefont {{Halbasch}},
  \citenamefont {{Zaletel}},\ and\ \citenamefont
  {{Swingle}}}]{2018JHEP...01..098N}%
  \BibitemOpen
  \bibfield  {author} {\bibinfo {author} {\bibfnamefont {P.}~\bibnamefont
  {{Nguyen}}}, \bibinfo {author} {\bibfnamefont {T.}~\bibnamefont {{Devakul}}},
  \bibinfo {author} {\bibfnamefont {M.~G.}\ \bibnamefont {{Halbasch}}},
  \bibinfo {author} {\bibfnamefont {M.~P.}\ \bibnamefont {{Zaletel}}}, \ and\
  \bibinfo {author} {\bibfnamefont {B.}~\bibnamefont {{Swingle}}},\ }\href
  {\doibase 10.1007/JHEP01(2018)098} {\bibfield  {journal} {\bibinfo  {journal}
  {Journal of High Energy Physics}\ }\textbf {\bibinfo {volume} {1}},\ \bibinfo
  {eid} {98} (\bibinfo {year} {2018})},\ \Eprint
  {http://arxiv.org/abs/1709.07424} {arXiv:1709.07424 [hep-th]} \BibitemShut
  {NoStop}%
\bibitem [{\citenamefont {{Umemoto}}\ and\ \citenamefont
  {{Zhou}}(2018)}]{2018arXiv180502625U}%
  \BibitemOpen
  \bibfield  {author} {\bibinfo {author} {\bibfnamefont {K.}~\bibnamefont
  {{Umemoto}}}\ and\ \bibinfo {author} {\bibfnamefont {Y.}~\bibnamefont
  {{Zhou}}},\ }\href@noop {} {\bibfield  {journal} {\bibinfo  {journal} {ArXiv
  e-prints}\ } (\bibinfo {year} {2018})},\ \Eprint
  {http://arxiv.org/abs/1805.02625} {arXiv:1805.02625 [hep-th]} \BibitemShut
  {NoStop}%
\bibitem [{\citenamefont {{Bao}}\ and\ \citenamefont
  {{Halpern}}(2018{\natexlab{a}})}]{2018JHEP...03..006B}%
  \BibitemOpen
  \bibfield  {author} {\bibinfo {author} {\bibfnamefont {N.}~\bibnamefont
  {{Bao}}}\ and\ \bibinfo {author} {\bibfnamefont {I.~F.}\ \bibnamefont
  {{Halpern}}},\ }\href {\doibase 10.1007/JHEP03(2018)006} {\bibfield
  {journal} {\bibinfo  {journal} {Journal of High Energy Physics}\ }\textbf
  {\bibinfo {volume} {3}},\ \bibinfo {eid} {6} (\bibinfo {year}
  {2018}{\natexlab{a}})},\ \Eprint {http://arxiv.org/abs/1710.07643}
  {arXiv:1710.07643 [hep-th]} \BibitemShut {NoStop}%
\bibitem [{\citenamefont {{Hirai}}\ \emph {et~al.}(2018)\citenamefont
  {{Hirai}}, \citenamefont {{Tamaoka}},\ and\ \citenamefont
  {{Yokoya}}}]{2018PTEP.2018f3B03H}%
  \BibitemOpen
  \bibfield  {author} {\bibinfo {author} {\bibfnamefont {H.}~\bibnamefont
  {{Hirai}}}, \bibinfo {author} {\bibfnamefont {K.}~\bibnamefont {{Tamaoka}}},
  \ and\ \bibinfo {author} {\bibfnamefont {T.}~\bibnamefont {{Yokoya}}},\
  }\href {\doibase 10.1093/ptep/pty063} {\bibfield  {journal} {\bibinfo
  {journal} {Progress of Theoretical and Experimental Physics}\ }\textbf
  {\bibinfo {volume} {2018}},\ \bibinfo {eid} {063B03} (\bibinfo {year}
  {2018})},\ \Eprint {http://arxiv.org/abs/1803.10539} {arXiv:1803.10539
  [hep-th]} \BibitemShut {NoStop}%
\bibitem [{\citenamefont {{Nomura}}\ \emph {et~al.}(2018)\citenamefont
  {{Nomura}}, \citenamefont {{Rath}},\ and\ \citenamefont
  {{Salzetta}}}]{2018PhRvD..98b6010N}%
  \BibitemOpen
  \bibfield  {author} {\bibinfo {author} {\bibfnamefont {Y.}~\bibnamefont
  {{Nomura}}}, \bibinfo {author} {\bibfnamefont {P.}~\bibnamefont {{Rath}}}, \
  and\ \bibinfo {author} {\bibfnamefont {N.}~\bibnamefont {{Salzetta}}},\
  }\href {\doibase 10.1103/PhysRevD.98.026010} {\bibfield  {journal} {\bibinfo
  {journal} {\prd}\ }\textbf {\bibinfo {volume} {98}},\ \bibinfo {eid} {026010}
  (\bibinfo {year} {2018})},\ \Eprint {http://arxiv.org/abs/1805.00523}
  {arXiv:1805.00523 [hep-th]} \BibitemShut {NoStop}%
\bibitem [{\citenamefont {{Bao}}\ and\ \citenamefont
  {{Halpern}}(2018{\natexlab{b}})}]{2018arXiv180500476B}%
  \BibitemOpen
  \bibfield  {author} {\bibinfo {author} {\bibfnamefont {N.}~\bibnamefont
  {{Bao}}}\ and\ \bibinfo {author} {\bibfnamefont {I.~F.}\ \bibnamefont
  {{Halpern}}},\ }\href@noop {} {\bibfield  {journal} {\bibinfo  {journal}
  {ArXiv e-prints}\ } (\bibinfo {year} {2018}{\natexlab{b}})},\ \Eprint
  {http://arxiv.org/abs/1805.00476} {arXiv:1805.00476 [hep-th]} \BibitemShut
  {NoStop}%
\bibitem [{\citenamefont {{Esp{\'{\i}}ndola}}\ \emph
  {et~al.}(2018)\citenamefont {{Esp{\'{\i}}ndola}}, \citenamefont {{Guijosa}},\
  and\ \citenamefont {{Pedraza}}}]{2018arXiv180405855E}%
  \BibitemOpen
  \bibfield  {author} {\bibinfo {author} {\bibfnamefont {R.}~\bibnamefont
  {{Esp{\'{\i}}ndola}}}, \bibinfo {author} {\bibfnamefont {A.}~\bibnamefont
  {{Guijosa}}}, \ and\ \bibinfo {author} {\bibfnamefont {J.~F.}\ \bibnamefont
  {{Pedraza}}},\ }\href@noop {} {\bibfield  {journal} {\bibinfo  {journal}
  {ArXiv e-prints}\ } (\bibinfo {year} {2018})},\ \Eprint
  {http://arxiv.org/abs/1804.05855} {arXiv:1804.05855 [hep-th]} \BibitemShut
  {NoStop}%
\bibitem [{\citenamefont {{Caputa}}\ \emph {et~al.}(2018)\citenamefont
  {{Caputa}}, \citenamefont {{Miyaji}}, \citenamefont {{Takayanagi}},\ and\
  \citenamefont {{Umemoto}}}]{2018arXiv181205268C}%
  \BibitemOpen
  \bibfield  {author} {\bibinfo {author} {\bibfnamefont {P.}~\bibnamefont
  {{Caputa}}}, \bibinfo {author} {\bibfnamefont {M.}~\bibnamefont {{Miyaji}}},
  \bibinfo {author} {\bibfnamefont {T.}~\bibnamefont {{Takayanagi}}}, \ and\
  \bibinfo {author} {\bibfnamefont {K.}~\bibnamefont {{Umemoto}}},\ }\href@noop
  {} {\bibfield  {journal} {\bibinfo  {journal} {arXiv e-prints}\ ,\ \bibinfo
  {eid} {arXiv:1812.05268}} (\bibinfo {year} {2018})},\ \Eprint
  {http://arxiv.org/abs/1812.05268} {arXiv:1812.05268 [hep-th]} \BibitemShut
  {NoStop}%
\bibitem [{\citenamefont {{Tamaoka}}(2019)}]{2019PhRvL.122n1601T}%
  \BibitemOpen
  \bibfield  {author} {\bibinfo {author} {\bibfnamefont {K.}~\bibnamefont
  {{Tamaoka}}},\ }\href {\doibase 10.1103/PhysRevLett.122.141601} {\bibfield
  {journal} {\bibinfo  {journal} {\prl}\ }\textbf {\bibinfo {volume} {122}},\
  \bibinfo {eid} {141601} (\bibinfo {year} {2019})},\ \Eprint
  {http://arxiv.org/abs/1809.09109} {arXiv:1809.09109 [hep-th]} \BibitemShut
  {NoStop}%
\bibitem [{\citenamefont {{Czech}}\ \emph {et~al.}(2015)\citenamefont
  {{Czech}}, \citenamefont {{Lamprou}}, \citenamefont {{McCandlish}},\ and\
  \citenamefont {{Sully}}}]{2015JHEP...10..175C}%
  \BibitemOpen
  \bibfield  {author} {\bibinfo {author} {\bibfnamefont {B.}~\bibnamefont
  {{Czech}}}, \bibinfo {author} {\bibfnamefont {L.}~\bibnamefont {{Lamprou}}},
  \bibinfo {author} {\bibfnamefont {S.}~\bibnamefont {{McCandlish}}}, \ and\
  \bibinfo {author} {\bibfnamefont {J.}~\bibnamefont {{Sully}}},\ }\href
  {\doibase 10.1007/JHEP10(2015)175} {\bibfield  {journal} {\bibinfo  {journal}
  {Journal of High Energy Physics}\ }\textbf {\bibinfo {volume} {2015}},\
  \bibinfo {eid} {175} (\bibinfo {year} {2015})},\ \Eprint
  {http://arxiv.org/abs/1505.05515} {arXiv:1505.05515 [hep-th]} \BibitemShut
  {NoStop}%
\bibitem [{\citenamefont {{Czech}}\ \emph {et~al.}(2016)\citenamefont
  {{Czech}}, \citenamefont {{Lamprou}}, \citenamefont {{McCandlish}},\ and\
  \citenamefont {{Sully}}}]{2016JHEP...07..100C}%
  \BibitemOpen
  \bibfield  {author} {\bibinfo {author} {\bibfnamefont {B.}~\bibnamefont
  {{Czech}}}, \bibinfo {author} {\bibfnamefont {L.}~\bibnamefont {{Lamprou}}},
  \bibinfo {author} {\bibfnamefont {S.}~\bibnamefont {{McCandlish}}}, \ and\
  \bibinfo {author} {\bibfnamefont {J.}~\bibnamefont {{Sully}}},\ }\href
  {\doibase 10.1007/JHEP07(2016)100} {\bibfield  {journal} {\bibinfo  {journal}
  {Journal of High Energy Physics}\ }\textbf {\bibinfo {volume} {2016}},\
  \bibinfo {eid} {100} (\bibinfo {year} {2016})},\ \Eprint
  {http://arxiv.org/abs/1512.01548} {arXiv:1512.01548 [hep-th]} \BibitemShut
  {NoStop}%
\bibitem [{\citenamefont {{Nozaki}}\ \emph
  {et~al.}(2013{\natexlab{a}})\citenamefont {{Nozaki}}, \citenamefont
  {{Numasawa}},\ and\ \citenamefont {{Takayanagi}}}]{2013JHEP...05..080N}%
  \BibitemOpen
  \bibfield  {author} {\bibinfo {author} {\bibfnamefont {M.}~\bibnamefont
  {{Nozaki}}}, \bibinfo {author} {\bibfnamefont {T.}~\bibnamefont
  {{Numasawa}}}, \ and\ \bibinfo {author} {\bibfnamefont {T.}~\bibnamefont
  {{Takayanagi}}},\ }\href {\doibase 10.1007/JHEP05(2013)080} {\bibfield
  {journal} {\bibinfo  {journal} {Journal of High Energy Physics}\ }\textbf
  {\bibinfo {volume} {2013}},\ \bibinfo {eid} {80} (\bibinfo {year}
  {2013}{\natexlab{a}})},\ \Eprint {http://arxiv.org/abs/1302.5703}
  {arXiv:1302.5703 [hep-th]} \BibitemShut {NoStop}%
\bibitem [{\citenamefont {{Chen}}\ \emph {et~al.}(2018)\citenamefont {{Chen}},
  \citenamefont {{Shu}},\ and\ \citenamefont {{Wu}}}]{2018arXiv180400441C}%
  \BibitemOpen
  \bibfield  {author} {\bibinfo {author} {\bibfnamefont {C.-B.}\ \bibnamefont
  {{Chen}}}, \bibinfo {author} {\bibfnamefont {F.-W.}\ \bibnamefont {{Shu}}}, \
  and\ \bibinfo {author} {\bibfnamefont {M.-H.}\ \bibnamefont {{Wu}}},\
  }\href@noop {} {\bibfield  {journal} {\bibinfo  {journal} {arXiv e-prints}\
  ,\ \bibinfo {eid} {arXiv:1804.00441}} (\bibinfo {year} {2018})},\ \Eprint
  {http://arxiv.org/abs/1804.00441} {arXiv:1804.00441 [hep-th]} \BibitemShut
  {NoStop}%
\bibitem [{\citenamefont {{Ugajin}}(2013)}]{2013arXiv1311.2562U}%
  \BibitemOpen
  \bibfield  {author} {\bibinfo {author} {\bibfnamefont {T.}~\bibnamefont
  {{Ugajin}}},\ }\href@noop {} {\bibfield  {journal} {\bibinfo  {journal}
  {ArXiv e-prints}\ } (\bibinfo {year} {2013})},\ \Eprint
  {http://arxiv.org/abs/1311.2562} {arXiv:1311.2562 [hep-th]} \BibitemShut
  {NoStop}%
\bibitem [{\citenamefont {{Calabrese}}\ and\ \citenamefont
  {{Cardy}}(2006)}]{2006PhRvL..96m6801C}%
  \BibitemOpen
  \bibfield  {author} {\bibinfo {author} {\bibfnamefont {P.}~\bibnamefont
  {{Calabrese}}}\ and\ \bibinfo {author} {\bibfnamefont {J.}~\bibnamefont
  {{Cardy}}},\ }\href {\doibase 10.1103/PhysRevLett.96.136801} {\bibfield
  {journal} {\bibinfo  {journal} {\prl}\ }\textbf {\bibinfo {volume} {96}},\
  \bibinfo {eid} {136801} (\bibinfo {year} {2006})},\ \Eprint
  {http://arxiv.org/abs/cond-mat/0601225} {arXiv:cond-mat/0601225
  [cond-mat.stat-mech]} \BibitemShut {NoStop}%
\bibitem [{\citenamefont {{Calabrese}}\ and\ \citenamefont
  {{Cardy}}(2007)}]{2007JSMTE..10....4C}%
  \BibitemOpen
  \bibfield  {author} {\bibinfo {author} {\bibfnamefont {P.}~\bibnamefont
  {{Calabrese}}}\ and\ \bibinfo {author} {\bibfnamefont {J.}~\bibnamefont
  {{Cardy}}},\ }\href {\doibase 10.1088/1742-5468/2007/10/P10004} {\bibfield
  {journal} {\bibinfo  {journal} {Journal of Statistical Mechanics: Theory and
  Experiment}\ }\textbf {\bibinfo {volume} {2007}},\ \bibinfo {pages} {10004}
  (\bibinfo {year} {2007})},\ \Eprint {http://arxiv.org/abs/0708.3750}
  {arXiv:0708.3750 [cond-mat.stat-mech]} \BibitemShut {NoStop}%
\bibitem [{\citenamefont {{Calabrese}}\ and\ \citenamefont
  {{Cardy}}(2005)}]{2005JSMTE..04..010C}%
  \BibitemOpen
  \bibfield  {author} {\bibinfo {author} {\bibfnamefont {P.}~\bibnamefont
  {{Calabrese}}}\ and\ \bibinfo {author} {\bibfnamefont {J.}~\bibnamefont
  {{Cardy}}},\ }\href {\doibase 10.1088/1742-5468/2005/04/P04010} {\bibfield
  {journal} {\bibinfo  {journal} {Journal of Statistical Mechanics: Theory and
  Experiment}\ }\textbf {\bibinfo {volume} {2005}},\ \bibinfo {pages} {04010}
  (\bibinfo {year} {2005})},\ \Eprint {http://arxiv.org/abs/cond-mat/0503393}
  {arXiv:cond-mat/0503393 [cond-mat.stat-mech]} \BibitemShut {NoStop}%
\bibitem [{\citenamefont {{Nozaki}}\ \emph
  {et~al.}(2013{\natexlab{b}})\citenamefont {{Nozaki}}, \citenamefont
  {{Numasawa}}, \citenamefont {{Prudenziati}},\ and\ \citenamefont
  {{Takayanagi}}}]{2013PhRvD..88b6012N}%
  \BibitemOpen
  \bibfield  {author} {\bibinfo {author} {\bibfnamefont {M.}~\bibnamefont
  {{Nozaki}}}, \bibinfo {author} {\bibfnamefont {T.}~\bibnamefont
  {{Numasawa}}}, \bibinfo {author} {\bibfnamefont {A.}~\bibnamefont
  {{Prudenziati}}}, \ and\ \bibinfo {author} {\bibfnamefont {T.}~\bibnamefont
  {{Takayanagi}}},\ }\href {\doibase 10.1103/PhysRevD.88.026012} {\bibfield
  {journal} {\bibinfo  {journal} {\prd}\ }\textbf {\bibinfo {volume} {88}},\
  \bibinfo {eid} {026012} (\bibinfo {year} {2013}{\natexlab{b}})},\ \Eprint
  {http://arxiv.org/abs/1304.7100} {arXiv:1304.7100 [hep-th]} \BibitemShut
  {NoStop}%
\bibitem [{\citenamefont {{Nahum}}\ \emph {et~al.}(2017)\citenamefont
  {{Nahum}}, \citenamefont {{Ruhman}}, \citenamefont {{Vijay}},\ and\
  \citenamefont {{Haah}}}]{2017PhRvX...7c1016N}%
  \BibitemOpen
  \bibfield  {author} {\bibinfo {author} {\bibfnamefont {A.}~\bibnamefont
  {{Nahum}}}, \bibinfo {author} {\bibfnamefont {J.}~\bibnamefont {{Ruhman}}},
  \bibinfo {author} {\bibfnamefont {S.}~\bibnamefont {{Vijay}}}, \ and\
  \bibinfo {author} {\bibfnamefont {J.}~\bibnamefont {{Haah}}},\ }\href
  {\doibase 10.1103/PhysRevX.7.031016} {\bibfield  {journal} {\bibinfo
  {journal} {Physical Review X}\ }\textbf {\bibinfo {volume} {7}},\ \bibinfo
  {eid} {031016} (\bibinfo {year} {2017})},\ \Eprint
  {http://arxiv.org/abs/1608.06950} {arXiv:1608.06950 [cond-mat.stat-mech]}
  \BibitemShut {NoStop}%
\bibitem [{\citenamefont {{Nahum}}\ \emph {et~al.}(2018)\citenamefont
  {{Nahum}}, \citenamefont {{Vijay}},\ and\ \citenamefont
  {{Haah}}}]{2018PhRvX...8b1014N}%
  \BibitemOpen
  \bibfield  {author} {\bibinfo {author} {\bibfnamefont {A.}~\bibnamefont
  {{Nahum}}}, \bibinfo {author} {\bibfnamefont {S.}~\bibnamefont {{Vijay}}}, \
  and\ \bibinfo {author} {\bibfnamefont {J.}~\bibnamefont {{Haah}}},\ }\href
  {\doibase 10.1103/PhysRevX.8.021014} {\bibfield  {journal} {\bibinfo
  {journal} {Physical Review X}\ }\textbf {\bibinfo {volume} {8}},\ \bibinfo
  {eid} {021014} (\bibinfo {year} {2018})},\ \Eprint
  {http://arxiv.org/abs/1705.08975} {arXiv:1705.08975 [cond-mat.str-el]}
  \BibitemShut {NoStop}%
\bibitem [{\citenamefont {{von Keyserlingk}}\ \emph {et~al.}(2018)\citenamefont
  {{von Keyserlingk}}, \citenamefont {{Rakovszky}}, \citenamefont
  {{Pollmann}},\ and\ \citenamefont {{Sondhi}}}]{2018PhRvX...8b1013V}%
  \BibitemOpen
  \bibfield  {author} {\bibinfo {author} {\bibfnamefont {C.~W.}\ \bibnamefont
  {{von Keyserlingk}}}, \bibinfo {author} {\bibfnamefont {T.}~\bibnamefont
  {{Rakovszky}}}, \bibinfo {author} {\bibfnamefont {F.}~\bibnamefont
  {{Pollmann}}}, \ and\ \bibinfo {author} {\bibfnamefont {S.~L.}\ \bibnamefont
  {{Sondhi}}},\ }\href {\doibase 10.1103/PhysRevX.8.021013} {\bibfield
  {journal} {\bibinfo  {journal} {Physical Review X}\ }\textbf {\bibinfo
  {volume} {8}},\ \bibinfo {eid} {021013} (\bibinfo {year} {2018})},\ \Eprint
  {http://arxiv.org/abs/1705.08910} {arXiv:1705.08910 [cond-mat.str-el]}
  \BibitemShut {NoStop}%
\bibitem [{\citenamefont {{Jonay}}\ \emph {et~al.}(2018)\citenamefont
  {{Jonay}}, \citenamefont {{Huse}},\ and\ \citenamefont
  {{Nahum}}}]{2018arXiv180300089J}%
  \BibitemOpen
  \bibfield  {author} {\bibinfo {author} {\bibfnamefont {C.}~\bibnamefont
  {{Jonay}}}, \bibinfo {author} {\bibfnamefont {D.~A.}\ \bibnamefont {{Huse}}},
  \ and\ \bibinfo {author} {\bibfnamefont {A.}~\bibnamefont {{Nahum}}},\
  }\href@noop {} {\bibfield  {journal} {\bibinfo  {journal} {arXiv e-prints}\
  ,\ \bibinfo {eid} {arXiv:1803.00089}} (\bibinfo {year} {2018})},\ \Eprint
  {http://arxiv.org/abs/1803.00089} {arXiv:1803.00089 [cond-mat.stat-mech]}
  \BibitemShut {NoStop}%
\bibitem [{\citenamefont {{Harper}}\ \emph {et~al.}(2018)\citenamefont
  {{Harper}}, \citenamefont {{Headrick}},\ and\ \citenamefont
  {{Rolph}}}]{2018arXiv180704294H}%
  \BibitemOpen
  \bibfield  {author} {\bibinfo {author} {\bibfnamefont {J.}~\bibnamefont
  {{Harper}}}, \bibinfo {author} {\bibfnamefont {M.}~\bibnamefont
  {{Headrick}}}, \ and\ \bibinfo {author} {\bibfnamefont {A.}~\bibnamefont
  {{Rolph}}},\ }\href@noop {} {\bibfield  {journal} {\bibinfo  {journal} {arXiv
  e-prints}\ ,\ \bibinfo {eid} {arXiv:1807.04294}} (\bibinfo {year} {2018})},\
  \Eprint {http://arxiv.org/abs/1807.04294} {arXiv:1807.04294 [hep-th]}
  \BibitemShut {NoStop}%
\bibitem [{\citenamefont {{Botero}}\ and\ \citenamefont
  {{Reznik}}(2004)}]{2004PhRvA..70e2329B}%
  \BibitemOpen
  \bibfield  {author} {\bibinfo {author} {\bibfnamefont {A.}~\bibnamefont
  {{Botero}}}\ and\ \bibinfo {author} {\bibfnamefont {B.}~\bibnamefont
  {{Reznik}}},\ }\href {\doibase 10.1103/PhysRevA.70.052329} {\bibfield
  {journal} {\bibinfo  {journal} {\pra}\ }\textbf {\bibinfo {volume} {70}},\
  \bibinfo {eid} {052329} (\bibinfo {year} {2004})},\ \Eprint
  {http://arxiv.org/abs/quant-ph/0403233} {arXiv:quant-ph/0403233 [quant-ph]}
  \BibitemShut {NoStop}%
\bibitem [{\citenamefont {{De Nobili}}(2016)}]{deNobili_thesis}%
  \BibitemOpen
  \bibfield  {author} {\bibinfo {author} {\bibfnamefont {C.}~\bibnamefont {{De
  Nobili}}},\ }\href@noop {} {\  (\bibinfo {year} {2016})}\BibitemShut
  {NoStop}%
\end{thebibliography}
\end{document}